%% file: SMP-12-019_temp.tex
\pdfoutput=1

\documentclass[11pt,twoside,a4paper,cmspaper,final,collab]{cms-tdr}
\def\svnVersion{187804}\def\cmsCernNo{2013-016}\def\cmsCernDate{\today}\def\cmsMessage{Submitted to the Journal of High Energy Physics}

\begin{document}\cmsNoteHeader{SMP-12-019}

\hyphenation{had-ron-i-za-tion}
\hyphenation{cal-or-i-me-ter}
\hyphenation{de-vices}

\newcommand{\WtoLN}    {\ensuremath{\PW\to\ell\cPgn}}
\newcommand{\WtoEN}    {\ensuremath{\PW\to\Pe\cPgn}}
\newcommand{\WtoMN}    {\ensuremath{\PW\to\Pgm\cPgn}}
\newcommand{\ZtoBB}    {\ensuremath{\Z\to\bbbar}}
\newcommand{\ZtoNN}    {\ensuremath{\Z\to\cPgn\bar{\cPgn}}}
\newcommand{\ZtoLL}    {\ensuremath{\Z\to\ell\ell}}
\newcommand{\ZtoMM}    {\ensuremath{\Z\to\MM}}
\newcommand{\ZtoEE}    {\ensuremath{\Z\to\EE}}
\newcommand{\WmnJ}     {\ensuremath{\PW(\Pgm\cPgn)+\text{jets}}}
\newcommand{\ZmmJ}     {\ensuremath{\Z(\Pgm\Pgm)+\text{jets}}}
\newcommand{\ZnnJ}     {\ensuremath{\Z(\cPgn\cPagn)+\text{jets}}}
\newcommand{\WJ}       {\ensuremath{\PW + \text{jets}}}
\newcommand{\HBB}      {\ensuremath{\PH\to\bbbar}}
\newcommand{\HTT}      {\ensuremath{\PH\to\TT}}
\newcommand{\mtW}      {\ensuremath{m_{\mathrm{T}}}}
\newcommand{\ptl}      {\ensuremath{p_{\mathrm{T}}^{\ell}}}
\newcommand{\MyZ}      {\ensuremath{\Z}}
\newcommand{\MyW}      {\ensuremath{\W}}
\newcommand{\MyH}      {\ensuremath{\PH}}
\newcommand{\Vudscg}   {\ensuremath{\mathrm{V+udscg}}}
\newcommand{\Wudscg}   {\ensuremath{\mathrm{W+udscg}}}
\newcommand{\Wenudscg} {\ensuremath{\PW(\Pe\cPgn)+\mathrm{udscg}}}
\newcommand{\Wmnudscg} {\ensuremath{\PW(\Pgm\cPgn)+\mathrm{udscg}}}
\newcommand{\Wenbb}    {\ensuremath{\PW(\Pe\cPgn)+\bbbar}}
\newcommand{\Wmnbb}    {\ensuremath{\PW(\Pgm\cPgn)+\bbbar}}
\newcommand{\Zeebb}    {\ensuremath{\Z(\Pe\Pe)+\bbbar}}
\newcommand{\Zmmbb}    {\ensuremath{\Z(\Pgm\Pgm)+\bbbar}}
\newcommand{\Zudsg}    {\ensuremath{\mathrm{Z+udsg}}}
\newcommand{\Zudscg}   {\ensuremath{\mathrm{Z+udscg}}}
\newcommand{\Zeeudscg} {\ensuremath{\Z(\Pe\Pe)+\mathrm{udscg}}}
\newcommand{\Zmmudscg} {\ensuremath{\Z(\Pgm\Pgm)+\mathrm{udscg}}}
\newcommand{\Zenbb}    {\ensuremath{\Z(\Pe\cPgn)+\bbbar}}
\newcommand{\Zmnbb}    {\ensuremath{\Z(\Pgm\cPgn)+\bbbar}}
\newcommand{\Wbb}      {\ensuremath{\mathrm{W\bbbar}}}
\newcommand{\Zbb}      {\ensuremath{\mathrm{Z\bbbar}}}
\newcommand{\Zcc}      {\ensuremath{\mathrm{Z\ccbar}}}
\newcommand{\Vbb}      {\ensuremath{\mathrm{V+\bbbar}}}
\newcommand{\Zll}      {\ensuremath{Z(\ell\ell)}}
\newcommand{\Zmm}      {\ensuremath{Z(\mu\mu)}}
\newcommand{\Zee}      {\ensuremath{Z(ee)}}
\newcommand{\Mjj}      {\ensuremath{M(\mathrm{jj})}}
\newcommand{\ptjj}     {\ensuremath{{\pt}(\mathrm{jj})}}
\newcommand{\MZ}       {\ensuremath{M_{\Z}}}
\newcommand{\dRJJ}     {\ensuremath{\Delta R(\mathrm{J1,J2})}}
\newcommand{\dEtaJJ}   {\ensuremath{\Delta \eta(\mathrm{J1,J2})}}
\newcommand{\dphiVH}   {\ensuremath{\Delta\phi(\mathrm{V,H})}}
\newcommand{\dphiWH}   {\ensuremath{\Delta\phi(\mathrm{W,H})}}
\newcommand{\dphiZH}   {\ensuremath{\Delta\phi(\mathrm{Z,H})}}
\newcommand{\dphiMJ}   {\ensuremath{\Delta\phi(\mathrm{pfMET,J})}}
\newcommand{\cosTH}    {\ensuremath{\cos{\theta^*}}}
\newcommand{\dThPull}  {\ensuremath{\Delta\theta_{\mathrm{pull}}}}
\newcommand{\ptV}      {\ensuremath{p_{\mathrm{T}}(\mathrm{V})}}
\newcommand{\ptH}      {\ensuremath{p_{\mathrm{T}}(\mathrm{H})}}
\newcommand{\ptZ}      {\ensuremath{p_{\mathrm{T}}(\Z)}}
\newcommand{\ptW}      {\ensuremath{p_{\mathrm{T}}(\W)}}
\newcommand{\Naj}      {\ensuremath{N_{\mathrm{aj}}}}
\newcommand{\Nal}      {\ensuremath{N_{\mathrm{al}}}}
\newcommand{\etaTF}    {\ensuremath{\abs{ \eta } < 2.5}}
\newcommand{\Bexp}     {\ensuremath{B_{\text{exp}}}}
\newcommand{\Bobs}     {\ensuremath{B_{\text{obs}}}}
\newcommand{\Nobs}     {\ensuremath{N_{\text{obs}}}}
\newcommand{\met}      {\ETmiss}

\renewcommand{\PYTHIA} {{\textsc{pythia6}}\xspace}
\newcommand{\PYTHIAEIGHT} {{\textsc{pythia8}}\xspace}
\renewcommand{\HERWIG} {{\textsc{herwig++}}\xspace}
\newcommand{\pdf}{\ensuremath{\mathrm{pdf}}}
\newcommand{\ta}{\Pgt\xspace}
\newcommand{\mtt}{\ensuremath{m_{\mathrm{t}\bar{\mathrm{t}}}}~}
\newcommand{\wboson}{\PW}
\newcommand{\zboson}{\cPZ}
\newcommand{\mzp}{\ensuremath{m_{\cPZpr}}}
\newcommand{\qq}{\cPq\cPaq\xspace}
\newcommand{\invpb}{\pbinv}
\newcommand{\intlumi}{5\fbinv}
\newcommand{\wmasssemilepdata}{ \ensuremath{m_\wboson^{\mathrm{DATA}} = 83.0 \pm 0.7 \;\GeVcc}}
\newcommand{\wmasssemilepmc}{ \ensuremath{m_\wboson^{\mathrm{MC}} = 82.5 \pm 0.3 \;\GeVcc}}
\newcommand{\topmasssemilepdata}{ \ensuremath{m_{\mathrm{t}}^{\mathrm{DATA}} = 177.1 \pm 2.0 \;\GeVcc}}
\newcommand{\topmasssemilepmc}{ \ensuremath{m_{\mathrm{t}}^{\mathrm{MC}} = 171.5 \pm 0.7 \;\GeVcc}}
\newcommand{\subjetjes}{\ensuremath{1.01 \pm 0.01}}
\newcommand{\mcuteffsemilepdata}{ \ensuremath{\epsilon_{m_\wboson}^{\mathrm{DATA}} = 0.49 \pm 0.01}}
\newcommand{\mcuteffsemilepmc}{ \ensuremath{\epsilon_{m_\wboson}^{\mathrm{MC}} =  0.50 \pm 0.01}}
\newcommand{\mucuteffsemilepdata}{ \ensuremath{\epsilon_\mu^{\mathrm{DATA}} = 0.64 \pm 0.01}}
\newcommand{\mucuteffsemilepmc}{ \ensuremath{\epsilon_\mu^{\mathrm{MC}} =  0.64 \pm 0.01}}
\newcommand{\scalefactor}{ \ensuremath{0.97 \pm 0.03}}
\newcommand{\scalefactorsqrd}{ \ensuremath{0.94 \pm 0.06}}
\newcommand{\topmasscuteffsemilepdata}{ \ensuremath{\epsilon_{m_t}^{\mathrm{DATA}} = 0.50 \pm 0.08}}
\newcommand{\topmasscuteffsemilepmc}{ \ensuremath{\epsilon_{m_t}^{\mathrm{MC}} =  0.65}}
\newcommand{\minmasscuteffsemilepdata}{ \ensuremath{\epsilon_{minMass}^{\mathrm{DATA}} = 0.71 \pm 0.16}}
\newcommand{\minmasscuteffsemilepmc}{ \ensuremath{\epsilon_{minMass}^{\mathrm{MC}} =  0.62}}
\newcommand{\scalefactorcatoptag}{ \ensuremath{0.89 \pm 0.28}}

\RCS$Revision: 185365 $
\RCS$HeadURL: svn+ssh://svn.cern.ch/reps/tdr2/papers/SMP-12-019/trunk/SMP-12-019.tex $
\RCS$Id: SMP-12-019.tex 185365 2013-05-09 16:10:23Z srappocc $
\newlength\cmsFigWidth
\ifthenelse{\boolean{cms@external}}{\setlength\cmsFigWidth{0.85\columnwidth}}{\setlength\cmsFigWidth{0.4\textwidth}}
\ifthenelse{\boolean{cms@external}}{\providecommand{\cmsLeft}{top}}{\providecommand{\cmsLeft}{left}}
\ifthenelse{\boolean{cms@external}}{\providecommand{\cmsRight}{bottom}}{\providecommand{\cmsRight}{right}}
\cmsNoteHeader{AN-12-137} 
\title{Studies of jet mass in dijet and W/Z$+$jet events}

\newcommand{\ifnpas}{\iffalse}
\newcommand{\ifpas}{\iftrue}

\date{\today}

\abstract{
Invariant mass spectra for jets reconstructed using the
anti-\kt and Cambridge--Aachen algorithms are studied for
different jet ``grooming'' techniques
in data corresponding to an integrated luminosity of
5\fbinv, recorded with the CMS detector in proton-proton collisions at
the LHC at a center-of-mass energy of 7\TeV.
Leading-order QCD predictions for inclusive dijet
and W/{Z}+jet production combined with parton-shower Monte Carlo
models are found to agree overall with
the data, and the agreement improves with the implementation
of jet grooming methods used to distinguish merged jets of large
transverse momentum from
softer QCD gluon radiation.
}

\hypersetup{%
pdfauthor={CMS Collaboration},%
pdftitle={Studies of jet mass in dijet and W/Z+jet events},%
pdfsubject={CMS},%
pdfkeywords={CMS, physics, software, computing}}

\maketitle 

\section{Introduction}
\label{sec:intro}
The variables most often used in analyses of jet production are jet directions and transverse momenta ($\pt$).
However, as jets are composite objects, their invariant masses ($m_J$) provide additional information that can be used to characterize their properties.
One motivation for investigating jet mass is that, at the Large Hadron Collider (LHC), massive standard model (SM) particles such as \PW\ and \Z bosons and top quarks are often produced with large Lorentz boosts, and, when such particles decay into quarks, the masses of the evolved jets can be used to discriminate them from lighter objects generated in quantum-chromodynamic (QCD) radiative processes. The same argument also holds for any new massive particles produced at the LHC.
For sufficiently large boosts, all the decay products tend to be emitted as collimated groupings into small sections of the detector, and the resulting particles can be clustered into a single jet. Jet ``grooming'' techniques are designed to separate such merged jets from background. These new techniques have been found to be very promising for identifying decays of highly-boosted \PW\ bosons and top quarks, and in searches for Higgs bosons and other massive particles~\cite{jetsub}.
The main advantage of these grooming techniques is their ability to distinguish high $\pt$ jets that arise from decays of massive, possibly new, particles.  In addition, their robust performance is valuable in the presence
of additional interactions in an event (pileup), which is likely to provide an even greater challenge to such analyses in future higher-luminosity runs at the LHC.

Only a few of these promising approaches have been studied in data at the Tevatron~\cite{cdfJS} or at the LHC~\cite{atlasJS}. To understand these techniques in the context of searches for new phenomena, the jet mass must be well-modeled through leading-order (LO) or next-to-leading-order (NLO) Monte Carlo (MC) simulations.
Much recent theoretical work in QCD has focused on the computation of jet mass, including
predictions using advances in an effective field theory of jets
(soft collinear effective theory, SCET)
\cite{KhelifaKerfa:2011zu,Hornig:2011tg,Li:2011hy,Bauer:2011uc,Chien:2010kc,Schwartz:2007ib,Fleming:2007qr,Dasgupta:2001sh,Bauer:2006mk,Bauer:2006qp,Hornig:2011iu,Kelley:2011ng,Jouttenus:2009ns,Ellis:2009wj,Ellis:2010rwa,Cheung:2009sg,Kelley:2010qs,Banfi:2010pa,Bauer:2001yt,Dasgupta:2012hg}.
Studies of the kind reported in the present analysis can provide an understanding of the extent to which MC simulations that match matrix-element partons with parton showers can model the observed internal jet structure. Results of these studies can also be used to compare data with theoretical computations of jet mass, and to provide benchmarks for the use of these algorithms in searches for highly-boosted Higgs bosons, or new objects beyond the SM, especially by investigating some of the background processes expected in such analyses.

We present a measurement of jet mass in a sample of dijet events, and the first study of such distributions in V+jet events, where V refers to a \PW\ or \Z boson. The data correspond to an integrated luminosity of $5.0 \pm 0.2\fbinv$, collected by the Compact Muon Solenoid (CMS) experiment at the LHC in pp interactions at a center-of-mass energy of 7\TeV.  The analysis of these two types of final states provides complementary information because of their different parton-flavor content, since the selected dijet events are dominated by gluon-initiated jets, and the V+jet events often contain quark-initiated jets. We focus on measuring the jet mass after applying several jet grooming techniques involving ``filtering''~\cite{boostedHiggs}, ``trimming''~\cite{trimming}, and ``pruning''~\cite{pruning,pruning2} of jets, as discussed in detail below.
This work also presents the first attempt to measure the mass of trimmed and pruned jets.

To study the dependence of the differential distributions in $m_J$ on jet $\pt$,
we measure the distributions in intervals of jet transverse momentum.
Formally, this
can be expressed in terms of a double-differential cross section for jet production
($d^2\sigma/d\pt dm_J$) that is examined as a function of $m_J$ for several nonoverlapping
intervals in $\pt$:

\begin{equation}
\label{eq:pdf_mjet_i}
\sigma = \int_{m_J} \int_{\pt} \frac{\rd^2 \sigma(m_J,\pt)}{\rd m_J \,\rd\pt} \,\rd\pt \,\rd m_J = \sum_i \int_{m_J} \frac{\rd\sigma_i(m_J)}{\rd m_J} \,\rd m_J = \sum_i \sigma_i,
\end{equation}

\noindent where $i=1,2,3,\ldots$ refers to the $i^{\text{th}}$ interval in $\pt$,
and the sum of contributions over all $i$ is equal to the total observed cross section
$\sum_i \sigma_i = \sigma$.
The differential probability density as a function of $m_J$ for
each $\pt$ interval can therefore be written as

\begin{equation}
\label{eq:pdf_mjet_simple}
\rho_i(m_J) = \frac{1}{\sigma_i} \times \frac{\rd\sigma_i}{\rd m_J},\; \text{with} \int \rho_i(m_J) \,\rd m_J = 1.
\end{equation}

The distributions in reconstructed jet mass of Eq.~(\ref{eq:pdf_mjet_simple})
include corrections used to unfold jets to the ``particle'' level;
the $\pt$ intervals are defined for ungroomed jets,
following energy corrections for the response of the detector.

For the dijet analysis, $\pt$ and $m_J$ correspond to
the average transverse momentum and average jet mass of
the two leading jets (\ie, of highest $\pt$): $\pt^\mathrm{AVG} = ({\pt}_1 + {\pt}_2) / 2$ and
$m_J^\mathrm{AVG} = ({m_J}_1 + {m_J}_2) / 2$.
For the V+jet analysis, we use the $m_J$ and $\pt$ of the leading jet.
Both quantities depend on the nature of the jet grooming algorithm, as discussed in
Section~\ref{sec:algos}.

This paper is organized as follows. To introduce the subject, we first discuss jet clustering algorithms in Section~\ref{sec:algos}, focusing mainly on grooming techniques. After a brief description of the CMS detector and the MC samples in Section~\ref{sec:cms_detector}, we provide information pertaining to the collected data and a description of event reconstruction in Section~\ref{sec:trigreco}. Selection of events is then described in Section~\ref{sec:evsel_paper}, and the effect of pileup on jet mass is investigated in Section~\ref{sec:pileup}. This is followed in Section~\ref{sec:systematics} by the correction and unfolding procedures that are applied to the $m_J$ spectra and their corresponding systematic uncertainties. In Sections~\ref{sec:dijetresults} and~\ref{sec:vjetresults}, we present the results of the dijet and V+jet analyses, respectively. Finally, observations and remarks on the presented results are summarized in Section~\ref{sec:summary}.

The distributions shown are also stored in HEPData format~\cite{hepdata}.
\section{Jet clustering algorithms and grooming techniques}
\label{sec:algos}

\subsection{Sequential jet clustering algorithms}

Jets are defined through sequential, iterative jet clustering
algorithms that combine four-vectors of input pairs of particles
until certain criteria are satisfied and jets are formed.
For the jet algorithms considered in this paper, for each pair of particles $i$ and $j$,
a ``distance'' metric between
the two particles ($d_{ij}$), and the so-called ``beam distance''
for each particle ($d_{iB}$), are computed:

\begin{align}
\label{eq:dij}
d_{ij} &= \min({\pt}_i^{2n},{\pt}_j^{2n}) \Delta R_{ij}^2 / R^2 \\
\label{eq:diB}
d_{iB} &= {\pt}_i^{2n},
\end{align}

where ${\pt}_i$ and ${\pt}_j$ are the transverse momenta of particles
$i$ and $j$, respectively, ``$\min$'' refers
to the lesser of the two $\pt$ values,
the integer $n$ depends on the specific jet algorithm, $\Delta R_{ij} = \sqrt{(\Delta y_{ij})^2 + (\Delta\phi_{ij})^2 }$
is the distance between $i$ and $j$ in
rapidity ($y = \frac{1}{2} \ln (E + p_{z})/(E - p_{z})$) and azimuth ($\phi$),
and $R$ is the ``size'' parameter of order unity~\cite{ktalg}, with all angles expressed in radians.
The particle pair $(i,j)$ with smallest $d_{ij}$ is combined into a
single object. All distances are recalculated using the new object, and the procedure is
repeated until, for a given object $i$, all the $d_{ij}$ are greater
than $d_{iB}$. Object $i$ is then
classified as a jet and not considered further in the algorithm. The process is
repeated until all input particles are clustered into jets.

The value for $n$ in Eqs.~(\ref{eq:dij}) and~(\ref{eq:diB}) governs
the topological properties of the jets.
For $n=1$ the procedure is referred to as the
$\kt$ algorithm (KT). The KT jets tend to have irregular
shapes and are especially useful for reconstructing jets of lower momentum~\cite{ktalg}.
For this reason, they are also sensitive to the presence of
low-$\pt$ pileup (PU) contributions, and are
used to compute the mean $\pt$ per unit area (in $(y,\phi)$) of an event~\cite{jetarea_fastjet}.
For $n=-1$, the procedure
is called the anti-$\kt$ (AK) algorithm, with features
close to an idealized cone algorithm.
The AK algorithm is used extensively
in LHC experiments and by the theoretical community for
finding well-separated jets~\cite{ktalg}. For $n=0$, the procedure
is called the Cambridge--Aachen (CA) algorithm. This relies only on angular
information, and, like the $\kt$ algorithm,
provides irregularly-shaped jets in $(y,\phi)$. The CA algorithm is useful in identifying
jet substructure~\cite{CAcambridge,CAaachen,Butterworth:2002tt}.

Jet grooming techniques~\cite{pruning} that reduce the impact of
contributions from the underlying event (UE),
PU, and low-$\pt$ gluon
radiation can be useful irrespective of the specific
nature of analysis.
These kinds of contributions to jets are typically soft and
diffuse, and hence contribute energy to the jet proportional to the
area~\cite{jetarea_fastjet}. Because grooming techniques reduce the areas
of jets without affecting the core components, the resulting jets are
less sensitive to
contributions from UE and PU, while still reflecting the kinematics of
the hard original process.
We consider three forms of grooming, referred to as
filtering, trimming, and pruning.
Such techniques can be applied to jets clustered through different algorithms (KT, AK, or CA).
For the dijet analysis, we choose to cluster jets with the anti-$\kt$
algorithm with $R=0.7$ (AK7), as these are used extensively at
CMS. For the V+jet analysis, in addition to AK7 jets,
we also study CA jets with $R=0.8$ (CA8), considered in
recent publications involving top-quark tagging~\cite{EXO-11-006},
and with $R=1.2$ (CA12), which was
proposed for analyses involving highly-boosted
objects~\cite{boostedHiggs}.
After the initial jet clustering with AK7, CA8, or CA12,
the constituents of those jets are reclustered with a (possibly different)
jet algorithm (\eg, KT, CA, or AK), applying additional grooming conditions
to the sequence of selection criteria used for clustering. The
optimal choice of this secondary clustering algorithm depends
on the grooming technique, as described below.
For the techniques we have investigated, the parameters chosen for the
algorithms correspond to those chosen by Refs.~\cite{boostedHiggs,trimming,pruning,pruning2}, nevertheless
specific optimization would appear to be advisable for all
well-defined searches for new phenomena.

\subsection{Filtering algorithm}

The ``mass-drop/filtering'' procedure aims to identify symmetric
splitting of jets of large $\pt$ that have large $m_J$ values. It was
proposed initially for use in searches for the Higgs
boson~\cite{boostedHiggs}, but we
consider just the filtering aspects of this algorithm for grooming jets.

For each jet obtained in the initial clustering procedure, the filtering algorithm defines
a new, groomed jet through the following algorithm:
(i) the constituents of each jet are reclustered using the CA algorithm
with $R=0.3$, thereby defining $n$ new subjets $s_1,\ldots,s_n$, ordered in
descending $\pt$, and
(ii) the four-momentum of the new jet is defined by the four-vector sum over
the three subjets of hardest $\pt$, or in the rare case that $n<3$,
just these remaining subjets define the new jet.

\noindent
The new jet has fewer particles than the initial jet, thereby reducing
the contribution from effects such as underlying event and pileup, and
the new $m_J$ and $\pt$ values are therefore smaller than those of the initial jet.
As will be demonstrated in Section~\ref{section:grommedjetmass}, with
this choice of parameters, filtering
removes the fewest jet constituents, and
is therefore the least aggressive of the investigated
 jet grooming techniques.

\subsection{Trimming algorithm}

Trimming ignores particles within a jet that fall below
a dynamic threshold in $\pt$~\cite{trimming}.
It reclusters the jet's constituents using the $\kt$
algorithm with a radius
$R_\text{sub}$, accepting only the subjets that have
$ {\pt}_\text{sub} > f_\text{cut} \lambda_\text{hard}$, where $f_\text{cut}$
is a dimensionless cutoff parameter, and $\lambda_\text{hard}$ is some
hard QCD scale chosen to equal the $\pt$ of the original jet.
The $R_\text{sub}$ and $f_\text{cut}$ parameters of the algorithm are
taken to be 0.2 and 0.03, respectively.
As will be demonstrated, with this choice of parameters, trimming
removes more jet constituents than the filtering procedure, but fewer jet
constituents than pruning, and corresponds therefore to
a moderately aggressive jet grooming technique.

\subsection{Pruning algorithm}

Following the clustering of jets using the original
algorithm (either AK7, CA8,
or CA12), the pruning algorithm~\cite{pruning,pruning2} reclusters the constituents
of the jet through the CA algorithm, using the same distance parameter, but
additional conditions beyond those given in Eq.~(\ref{eq:dij}). In particular,
the softer of the two particles $i$ and $j$ to be merged is removed when
the following conditions are met:
\begin{align}
z_{ij} & =  \frac{\min({\pt}_i,{\pt}_j)}{{\pt}_i + {\pt}_j} < z_{\text{cut}} \\
\Delta R_{ij} & >  D_{\text{cut}} \equiv \alpha \cdot \frac{m_J}{\pt},
\end{align}
where
$m_J$ and $\pt$ are the mass and transverse momentum of the originally-clustered jet,
and $z_{\text{cut}}$ and $\alpha$ are parameters of the algorithm,
chosen to be 0.1 and 0.5, respectively.
In our particular choice of parameters, we have chosen to
divide the jet into two ``exclusive'' subjets (similarly
to the exclusive $\kt$ algorithm~\cite{ktalg}, where one clusters
constituents until the jets are all separated by the parameter
$R$ in Eq.~\ref{eq:dij}).
As will be demonstrated, with this choice of parameters, pruning
removes the largest number of jet constituents, and can
therefore be regarded as
the most aggressive jet grooming technique investigated.
It was previously used in the CMS search for
$\ttbar$ resonances~\cite{EXO-11-006}.

\subsection{Groomed jet mass}
\label{section:grommedjetmass}

Figure~\ref{figs:histAK7PtAvgVsMjetGroomOverReco_ratioPlots}
shows a comparison of distributions in the dijet sample for the ratio of groomed AK7 jet mass
to the mass of the matched ungroomed AK7 jet, for our
three grooming techniques, for data and for \PYTHIA MC
simulation~\cite{pythia}, using the Z2 tune.
Three distributions are shown for each grooming technique:
(i) the reconstructed data (``data RECO''), (ii)
the reconstructed simulated \PYTHIA data (``PYTHIA RECO''), and
(iii) the generated particle-level jets from \PYTHIA (``PYTHIA GEN'').
These three grooming techniques
involve different jet algorithms for grooming
(CA for filtering and pruning, $\kt$ for trimming)
once the jets are found with AK7.
The data and the simulation exhibit similar behavior. In general,
the filtering algorithm is the least aggressive grooming technique,
with groomed jet masses close to the ungroomed values.
The trimming algorithm is moderately aggressive, and the
pruning algorithm is the most aggressive of the three. With pruning, a bimodal
distribution begins to appear,
which is typical of our implementation of this algorithm
as we require clustering into two exclusive subjets. In cases where
the pruned jet mass is small,
jets usually have most of their energy configured in ``core'' components,
with little gluon radiation, which leads to narrow jets. When the pruned jet
mass is large, the jets are split more symmetrically,
which can be realized in events with gluons splitting
into two nodes that fall within
 $\Delta R=0.7$ of the original parton.

\begin{figure}[htbp]
\centering
\includegraphics[width=0.95\textwidth]{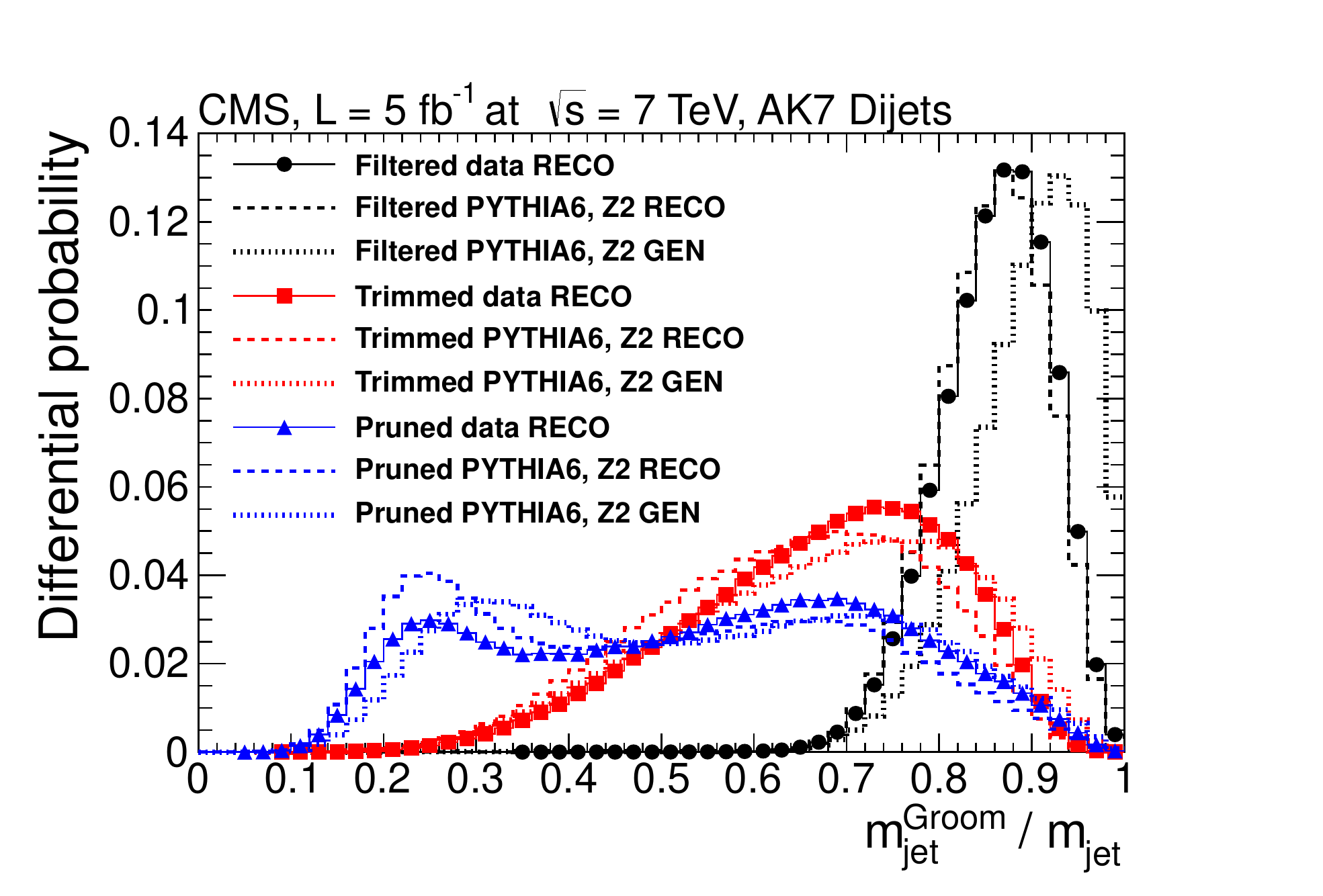}
\caption{Distributions in differential probability for ratios of the jet mass of groomed jets
to their
corresponding ungroomed values, for both dijet data and \PYTHIA (tune Z2) MC
simulation, for the
three grooming techniques discussed in the text:
(i) filtering (circles, peaking near $0.9$),
(ii) trimming (squares, peaking near $0.75$), and
(iii) pruning (triangles, more dispersed).
\label{figs:histAK7PtAvgVsMjetGroomOverReco_ratioPlots}}
\end{figure}

\section{The CMS detector and simulation}
\label{sec:cms_detector}

The CMS detector~\cite{:2008zzk}
is a general-purpose device with
many features suited for reconstruction of
energetic jets, specifically, the finely segmented electromagnetic
and hadronic calorimeters and charged-particle tracking detectors.

CMS uses a right-handed coordinate system, with origin
defined by the center of the CMS detector,
the $x$ axis pointing to the center of the LHC ring,
the $y$ axis pointing up, perpendicular to the plane of the LHC ring,
and the $z$ axis along the direction of the counterclockwise beam.
The polar angle $\theta$ is measured relative to the positive $z$
axis and the azimuthal angle $\phi$ relative to the $x$ axis in the $x$-$y$ plane.

Charged particles are reconstructed in the inner silicon tracker,
which is immersed in a 3.8\unit{T} axial magnetic field.
The CMS tracking detector consists of an inner silicon pixel detector
composed of three concentric central layers and two sets of disks
arranged forward and backward of the center,
and up to ten silicon strip central layers and three inner and nine
outer strip disks forward and backward of the center.
This arrangement provides
full azimuthal coverage for $\abs{\eta} < 2.5$, where
$\eta = -\ln[\tan(\theta/2)]$ is the pseudorapidity.
The pseudorapidity approximates the rapidity $y$
and
equals $y$ for massless particles.
Since many of the reconstructed jets
are not massless, we use the rapidity $y$ for characterizing
jets in this analysis.

A lead tungstate crystal electromagnetic calorimeter (ECAL) and
a brass/scintillator hadronic calorimeter (HCAL) surround the tracking
volume and provide photon, electron, and jet reconstruction up to $\abs{\eta}=3$.
The ECAL and HCAL cells are grouped into towers projecting radially
outward from the center of the detector.
In the central region ($\abs{\eta}<1.74$),
the towers have dimensions of $\Delta\eta = \Delta\phi = 0.087$
that increase at larger $\abs{\eta}$.
ECAL and HCAL cell energies above some chosen noise-suppression
thresholds are combined within each tower to define the tower energy.
Muons are measured in gas-ionization detectors embedded in the steel
return yoke outside the solenoid.
To improve reconstruction of jets, the tracking and calorimeter
information is combined in a ``particle-flow'' (PF)
algorithm~\cite{particleflow}, which is described in
Section~\ref{sec:reconstruction}.

For the analysis of dijet events, samples are simulated with
\PYTHIA.4 (Tune Z2) ~\cite{pythia,Field:2010bc},
\PYTHIAEIGHT  (Tune 4c)~\cite{pythia8},
and \HERWIG (Tune 23)~\cite{herwig}, and
propagated through the simulation of the CMS
detector based on \GEANTfour \cite{Geant4}.
Underlying event (UE) and pileup (PU) are included in the
simulations, which are also reweighted to have the simulated
PU distribution match the observed PU distribution
in the data.

For the V+jet analysis, events with a vector boson produced in
association with jets are simulated using \MADGRAPH 5.1~\cite{madgraph}.
This matrix element generator is also used to simulate $\ttbar$ events.
The \MADGRAPH events are subsequently subjected to parton showering, simulated
with \PYTHIA using the Z2 Tune~\cite{Field:2010bc}.
To compare hadronization in different generators, we generate V+jet
samples in which parton showering and hadronization are
simulated with \HERWIG.
Diboson (WW, WZ, and ZZ) events are also generated with \PYTHIA.
Single-top-quark samples are produced with \POWHEG~\cite{powheg}, and the
lepton enriched dijet samples are produced with \PYTHIA using the Z2 Tune.
CTEQ6L1~\cite{cteq} is the default set of parton distribution
functions used
in all these samples, except for the single-top-quark MC, which uses CTEQ6M.

\section{Triggers and event reconstruction}
\label{sec:trigreco}
\subsection{Dijet trigger selection}
\label{sec:dataSampleAndEventSelection}

Events are collected using single-jet triggers, which are
based on jets reconstructed only from calorimetric information.
This procedure yields inferior resolution to jets reconstructed
offline with PF constituents, but provides faster
reconstruction that meets trigger requirements.
As the instantaneous luminosity is time-dependent, the specific
jet-$\pt$ thresholds change with time.
The triggers used to select dijet events have partial overlap.
Those with lower-$\pt$ thresholds have high prescale settings to accommodate the higher
data-acquisition rates, and some
events selected with these lower-$\pt$ triggers are also collected at
higher thresholds.

To avoid double counting of phase space, each event is assigned
to a specific trigger.
To do this, we compute the trigger
efficiency as a function of reconstructed $\pt^\mathrm{AVG}$, select
an interval in trigger efficiency where the efficiency is maximum (${>}95$\%) for
that range of $\pt^\mathrm{AVG}$, and assign that trigger to the appropriate $\pt^\mathrm{AVG}$ interval.
The assignment is based on the
jet $\pt$ values
reconstructed offline (but not groomed).
Table~\ref{TriggerTurnOns}
shows the $\pt$ thresholds for each of the jet triggers used in the
analysis, and the corresponding intervals of $\pt$ to which the
triggered events are assigned.

\begin{table}[h]
  \centering
  \topcaption{Trigger $\pt$ thresholds for individual jets,
    and corresponding $\pt^\mathrm{AVG}$ intervals used to assign the
    triggered events in the dijet analysis.\label{TriggerTurnOns}}
  \begin{tabular}{ c|c}
    \hline
\rule{0pt}{12pt}
Trigger $\pt$ threshold (\GeVns) & $\pt^\mathrm{AVG}$ range (\GeVns{}) \\
\hline
190& 220--300  \\
240& 300--450  \\
370& $>$450 \\
   \hline
  \end{tabular}
\end{table}

\subsection{V+jet trigger selection}
\label{sec:dataSampleAndEventSelectionVjet}

Several triggers are also used to collect events corresponding to
the topology of V+jet events, where the V decays via electrons or
muons in the final state.
For the \PW+jet channels, the triggers consist of several single-lepton
triggers, with lepton identification criteria applied online.
To assure an acceptable event rate, leptons are required to be isolated from other
tracks and energy depositions in the calorimeters.
For the \PW$(\mu\nu_\mu)$ channel, the trigger thresholds for the
muon $\pt$ are in the range of 17 to 40\GeV.
The higher thresholds are used at higher instantaneous luminosity.
The combined trigger efficiency for signal events
that pass offline requirements
(described in Section~\ref{sec:evsel_paper}) is ${\approx} 92\%$.

For the $\PW(\Pe\Pgne)$ events, the electron $\pt$ threshold ranges
from 25 to 65\GeV.
To enhance the fraction of \PW+jet events in the data, the
single-electron triggers are also required to have minimum thresholds
on the magnitude of the imbalance
in transverse energy ($\met$) and on the transverse mass
($m_\mathrm{T}$) of the (electron + $\met$) system,
where $m_\mathrm{T}^2 = 2E_\mathrm{T}^{\Pe}\met(1-\cos\phi)$,
and $\phi$ is the angle between the directions of $\pt^{\Pe}$ and $\met$.
The combined efficiency for electron \PW+jet events that pass the offline
criteria is ${\approx} 99\%$.

 The $\Z(\mu\mu)$ channel uses the same single-muon triggers as the
 $\PW(\mu\nu_\mu)$ channel. The $\Z (\Pe\Pe)$ channel uses dielectron triggers
 with lower thresholds for $\pt$ (17 and 8\GeV), and additional isolation
 requirements. These triggers are 99\% efficient for all
 {\Z}+jet events that pass the final offline selection criteria.

\subsection{Binning jets as a function of \texorpdfstring{$\pt$}{pt}}
\label{sec:ptBinAssignment}

The jet $\pt$ bins introduced in Eq.~(\ref{eq:pdf_mjet_i}) are given in
Table~\ref{tab:ptBins} for V+jet and dijet events. The jet $\pt$ is re-evaluated for each grooming
algorithm.
Because there are large biases due to jet misassignment in the dijet
events, especially at small $\pt$ (when three particle-level jets are
often reconstructed as two jets in the detector, or vice versa),
the $\pt$ intervals for these events begin at 220\GeV.
Furthermore, the smaller number of events in the V+jet samples precludes the
study of these events beyond $\pt= 450\GeV$.

\begin{table}[h]
  \centering
  \topcaption{Intervals in ungroomed jet $\pt$ for the V+jet and dijet analyses. \label{tab:ptBins}}
  \begin{tabular}{ ccc}
    \hline
\rule{0pt}{12pt}
    Bin & $\pt$ interval (\GeVns{}) & Analysis\\
    \hline
    1 & 125--150 & V+jet \\
    2 & 150--220 & V+jet  \\
    3 & 220--300 & V+jet,dijet  \\
    4 & 300--450 & V+jet,dijet  \\
    5 & 450--500 & dijet  \\
    6 & 500--600 & dijet  \\
    7 & 600--800 & dijet  \\
    8 & 800--1000 & dijet  \\
    9& 1000--1500 & dijet  \\
   \hline
  \end{tabular}
\end{table}

\subsection{Event reconstruction}
\label{evrecosection}

\label{sec:preselection}
\label{sec:reconstruction}
As indicated above, events are reconstructed using the particle-flow
algorithm, which
combines the information from all subdetectors to reconstruct
the particle candidates
in an event.
The algorithm categorizes particles into muons,
electrons, photons, charged hadrons, and neutral hadrons.
The resulting PF candidates are passed through each jet clustering
algorithm of Section~\ref{sec:algos},
as implemented in \textsc{FastJet} (Version 3.0.1) \cite{fastjet1,fastjet2}.

The reconstructed interaction vertex characterized by the largest value
of $\sum_i ({\pt}^{\mathrm{trk}}_i)^2$, where ${\pt}^{\mathrm{trk}}_i$ is the transverse momentum of the
$i^{\mathrm{th}}$ charged track associated with the vertex, is defined
as the leading primary vertex (PV) of the event.
This vertex is used as the reference vertex for all PF objects in the event.
A pileup interaction can affect the reconstruction of
jet momenta and $\met$, as well as lepton isolation and b-tagging efficiency.
To mitigate these effects, a track-based algorithm is used to remove
all charged hadrons that are not consistent with originating from the
leading PV.

Electron reconstruction requires the matching of an energy cluster in the
ECAL with a track extrapolated from the silicon
tracker~\cite{CMS-PAS-EGM-10-004}.
Identification criteria based on the energy distribution of showers in the
ECAL
and consistency of tracks
with the primary vertex are imposed on electron candidates.
Additional requirements remove any electrons produced through
conversions of photons in detector material.
The analysis considers electrons only in the range of $\abs{\eta}<2.5$,
excluding the transition region $1.44<\abs{\eta}<1.57$ between the central and endcap ECAL detectors
because of poorer resolution for electrons in this region.
Muons are reconstructed using two algorithms~\cite{CMS-PAS-MUO-10-004}:
(i)~in which tracks in the silicon tracker are matched to signals in
the muon chambers, and (ii)~in which a global fit is performed to a track
seeded by signals in the external muon system.
The muon candidates are required to be reconstructed through both algorithms.
Additional identification criteria are imposed on muon candidates to reduce
the fraction of tracks misidentified as muons, and to reduce
contamination from muon
decays in flight.
These criteria include the number of hits detected
in the tracker and in the outer muon system, the quality of the fit to
a muon track, and its consistency of originating from the leading PV.

Charged leptons from V-boson decays are expected to be isolated from
other energy depositions in the event.
For each lepton candidate, a cone with radius 0.3 for muons and 0.4
for electrons is chosen around the direction of the
track at the event vertex.
When the scalar sum of the transverse momenta of reconstructed particles
within that cone, excluding the contribution from the
lepton candidate,
exceeds ${\approx} 10\%$ of the $\pt$ of the
lepton candidate, that lepton is ignored.
The exact isolation requirement depends on the $\eta$, $\pt$, and flavor of
the lepton.
Muons and electrons are required to have $\pt > 30$\GeV and $>80$\GeV, respectively.
The large threshold for electrons ensures good trigger efficiency.
To avoid double counting, isolated charged leptons are removed from
the list of PF objects that are clustered into jets.

After removal of isolated leptons and charged hadrons
from pileup vertices, only the
neutral hadron component from pileup remains and is included in the
jet clustering. 
This remaining component of pileup to the jet energy is removed by
applying a
correction based on a
mean $\pt$ per unit area of ($\Delta y \times \Delta \phi$)
originating
from neutral particles~\cite{jetarea_fastjet,jetarea_fastjet_pu}.
This quantity is computed using
the $\kt$
algorithm, and corrects the jet energy
by the amount of energy expected from pileup in the jet cone.
This ``active area'' method adds a large number of
soft ``ghost'' particles to the
clustering sequence to determine the effective area subtended by each
jet.
This procedure is done for all grooming algorithms just
as for the ungroomed jets.
The active area of a groomed jet is smaller than that of an ungroomed jet, and the pileup correction is therefore also smaller.
Different responses in the endcap and central barrel calorimeters
necessitate using $\eta$-dependent jet corrections.
The amount of energy expected from the remnants of the hard collision
(the underlying
event) is estimated from minimum-bias data and MC events, and
is added back into the jet.

In addition, the pileup-subtracted jet four-momenta in data are
corrected for nonlinearities in $\eta$ and $\pt$ by using a
$\pt$- and $\eta$-dependent correction to account for the difference
between the response in MC-simulated events and
data~\cite{citeJEC}.
The jet corrections are derived for the ungroomed
jet algorithms but are also applied to the groomed algorithms, thereby adding
additional systematic uncertainty in the energy of groomed jets.

\section{Event selection}
\label{sec:evsel_paper}

We apply several other selection criteria to minimize instrumental
background and electronic noise. In particular,
accepted events must have at least one good primary vertex
(Section~\ref{sec:reconstruction}).
Backgrounds from additional beam interactions
are reduced by applying a variety of requirements on charged tracks.
Finally, calorimeter noise is minimized through restrictions on
timing and electronic pulse shapes expected for signals.

Dijet events are required to have at least two AK7 jets, each with
$\pt > 50$\GeV and $|y| < 2.5$,
and each jet must satisfy the jet quality criteria discussed in
Ref.~\cite{particleflow}.
No third-jet veto is applied.

Reconstruction of \PW\ and \Z\ bosons in V+jet events
begins with identification of charged leptons and a
calculation of $\met$, described in the previous
section.
Candidates for $\Z\to\ell^+\ell^-$ ($\ell=\Pe$ or $\mu$) decays are reconstructed by combining
two isolated electrons or muons and requiring the dilepton invariant
mass to be in the $80<M_{\ell\ell}<100\GeV$ range.
An accurate measurement of $\met$ is essential for distinguishing
the W signal from background processes.
The $\met$ in the event is defined using the PF objects,
and this analysis requires $\met > 50\GeV$.
Candidate $\WtoLN_\ell$ decays are identified primarily through the presence of
a significant $\met$ and a single isolated lepton of large $\pt$, with
\pt and \mtW\ of the \PW\ candidate obtained by combining the lepton
and the $\met$ vectors.

The analysis of V+jet events is mainly of interest for the
regime of $\pt^{\mathrm{V}}> 120$\GeV, in which the opposing jet tends to have large $\pt$ as well, because of momentum conservation.
In fact, the leading jet in each event (independent of clustering
algorithm and jet radius) is required to have $\pt > 125\GeV$ and $|y| < 2.5$.
A back-to-back topology between the vector boson and the leading jet
is ensured by the additional selection of $\Delta \phi(\text{V, jet})>2$ and
$\Delta R (\ell,\text{jet})>1$.
Requiring such highly boosted jets, in addition to the tight isolation
criteria for the leptons, greatly suppresses the background from multijet production.
In the $\WtoLN_\ell$+jet analysis, additional rejection
of multijet background is achieved by requiring
$m_\mathrm{T}(\PW) > 50\GeV$.
No subleading-jet veto is applied.

\begin{figure}[htbp]
\centering
\includegraphics[width=0.495\textwidth]{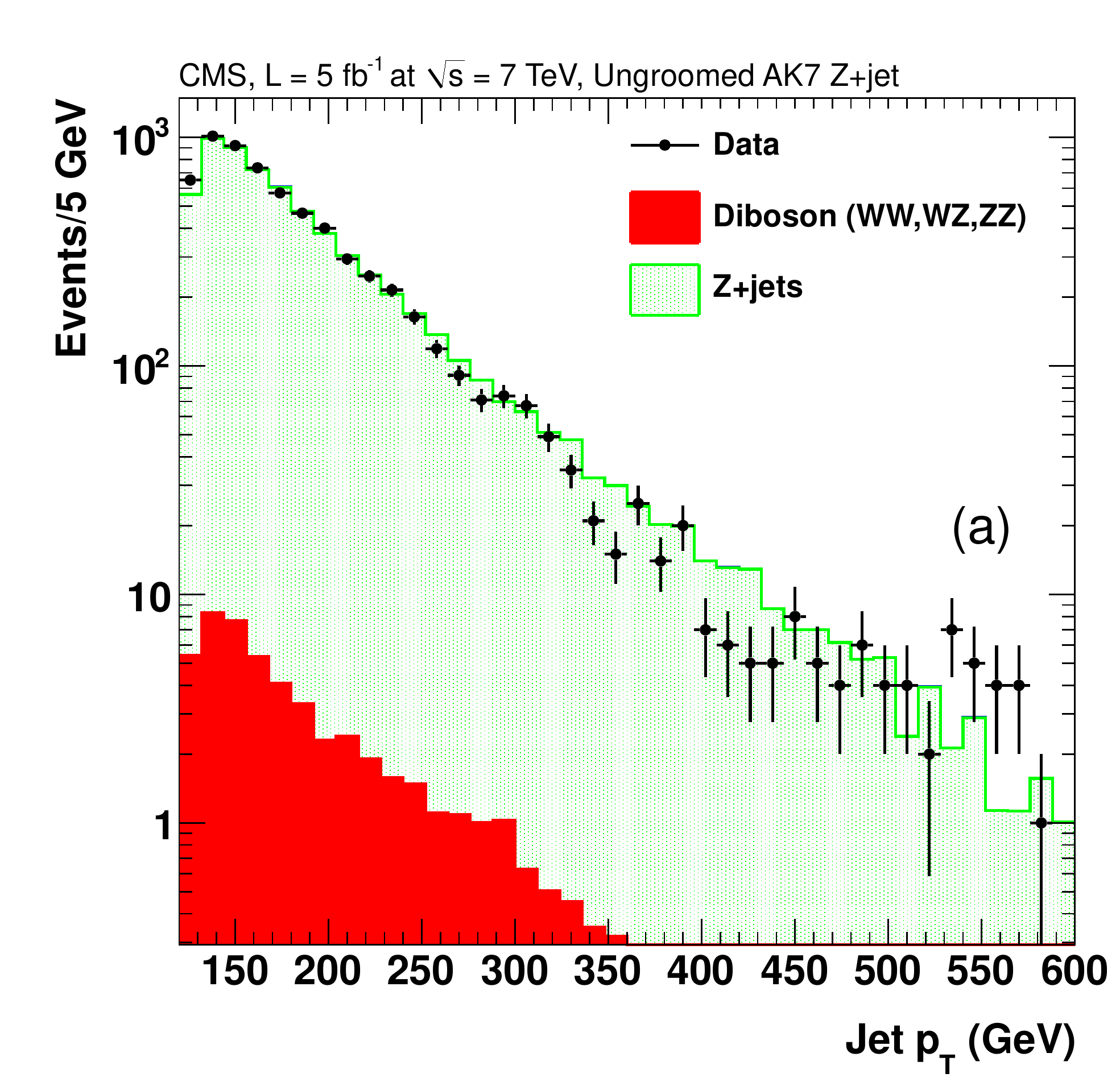}
\includegraphics[width=0.495\textwidth]{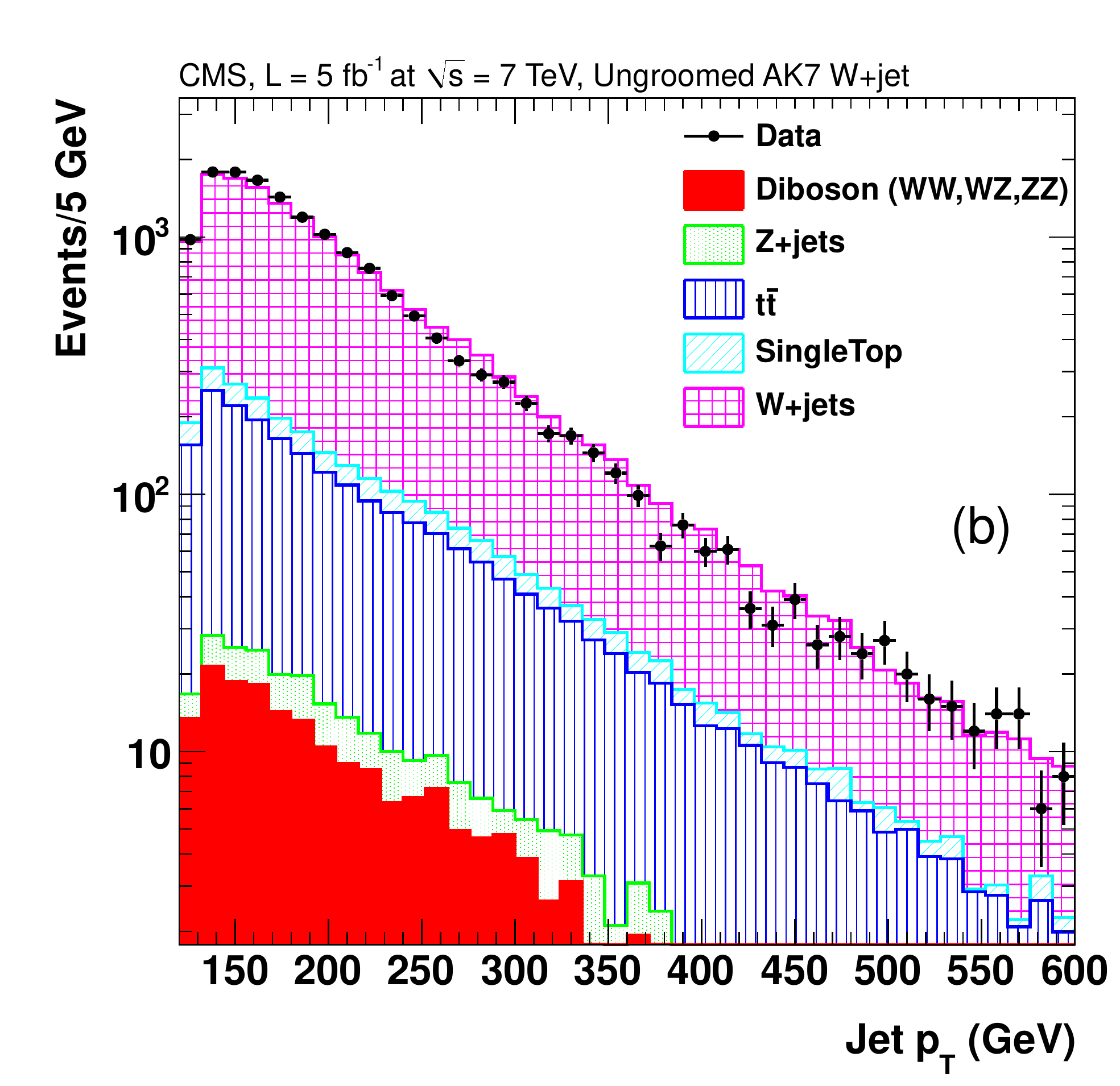}
\caption{The $\pt$ distribution for the leading AK7 jet in accepted (a)~\cPZ+jet and (b)~\PW+jet events.
\label{fig:Vjetpt}}
\end{figure}

Figures~\ref{fig:Vjetpt}(a) and (b) show the $\pt$ distributions
for the leading AK7 jet selected in Z+jet and W+jet candidate events,
respectively.
Given the unique signature for highly-boosted vector
bosons recoiling from jets, the selections suffice to define very pure samples of V+jet events.
In the $\Z(\ell\ell)$+jet analysis, the additional
constraint on dilepton mass removes almost all
background contributions, yielding a purity of ${\approx} 99\%$ for {\Z}+jet events, with
${\approx} 1\%$ contamination from diboson production.
The \PW+jet candidate sample contains ${\approx} 82\%$ \PW+jet events,
with small background contributions from $\ttbar$ (13\%),
single top-quark (3\%), and diboson and {\Z}+jet (1\% each) events based on MC simulation.
The small number of events expected from these processes are
subtracted using MC predictions for the jet mass
from the \PW+jet candidate events,
before correcting the data for detector effects.
Similarly, the small number of events expected from diboson production
are subtracted from the \Z+jet candidates.

\section{Influence of pileup on jet grooming algorithms}
\label{sec:pileup}

During the data taking the instantaneous LHC luminosity exceeded
${\approx}3.0 \times 10^{33}\percms$,
or an average of ten interactions per bunch crossing.
Such pileup collisions are not correlated with the hard-scattering
 process that triggers an interesting event, but present a background
 from low-$\pt$ interactions that can affect the measured energies of
jets and their observed masses.
 Methods to mitigate these effects are part of standard event
reconstruction, as discussed in Section~\ref{evrecosection},
and are essential for extracting correct jet multiplicities and energies.
The jet mass is expected to be particularly sensitive to pileup~\cite{jetsub}
for jets of large angular extent that contain many
particles. Grooming techniques are designed to reduce the effective
area of such jets and thereby minimize sensitivity to pileup.
We examine this issue through
studies of jet mass in the presence of pileup.

\begin{figure}[htbp]
\centering
\includegraphics[width=0.495\textwidth]{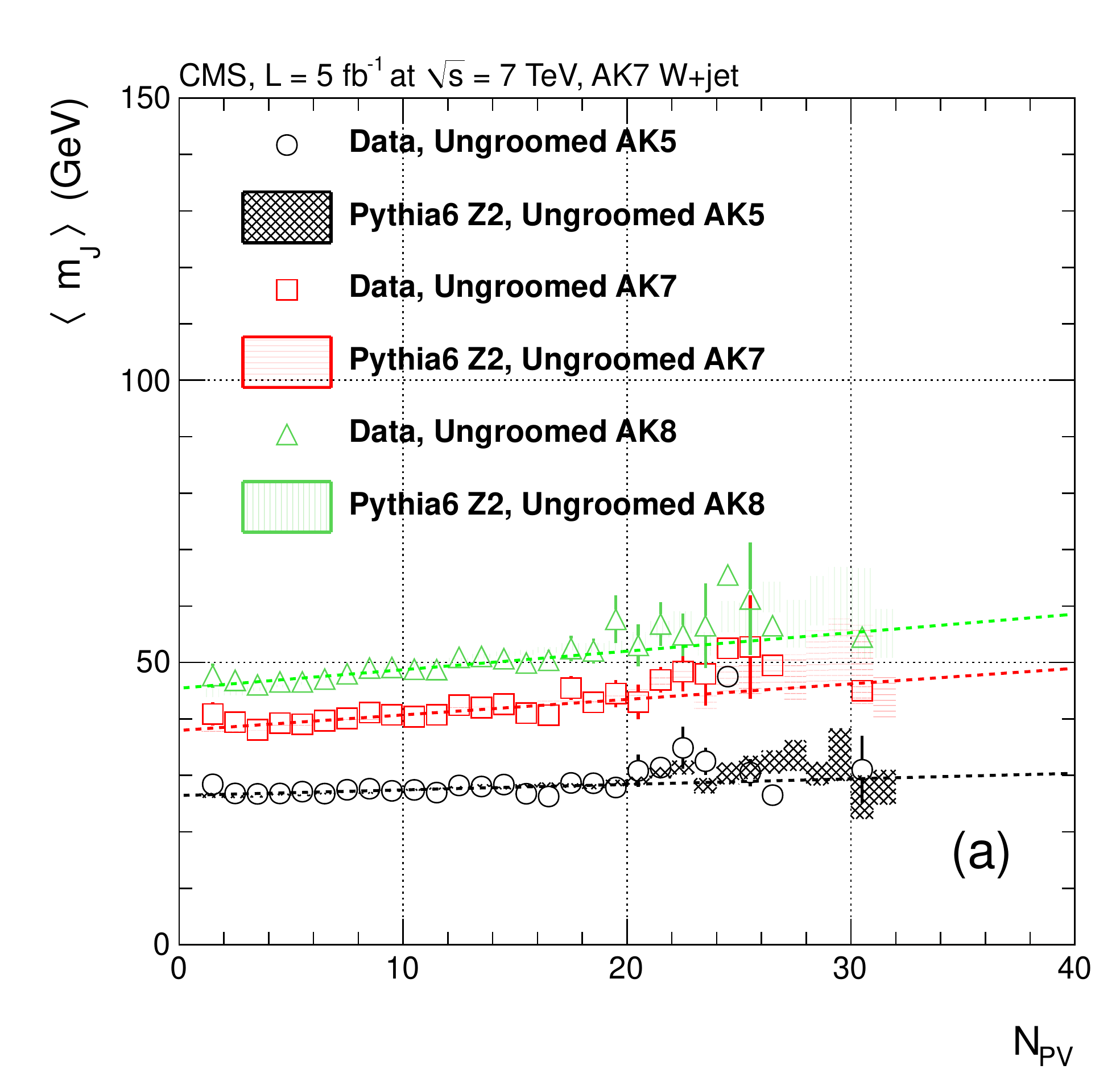}
\includegraphics[width=0.495\textwidth]{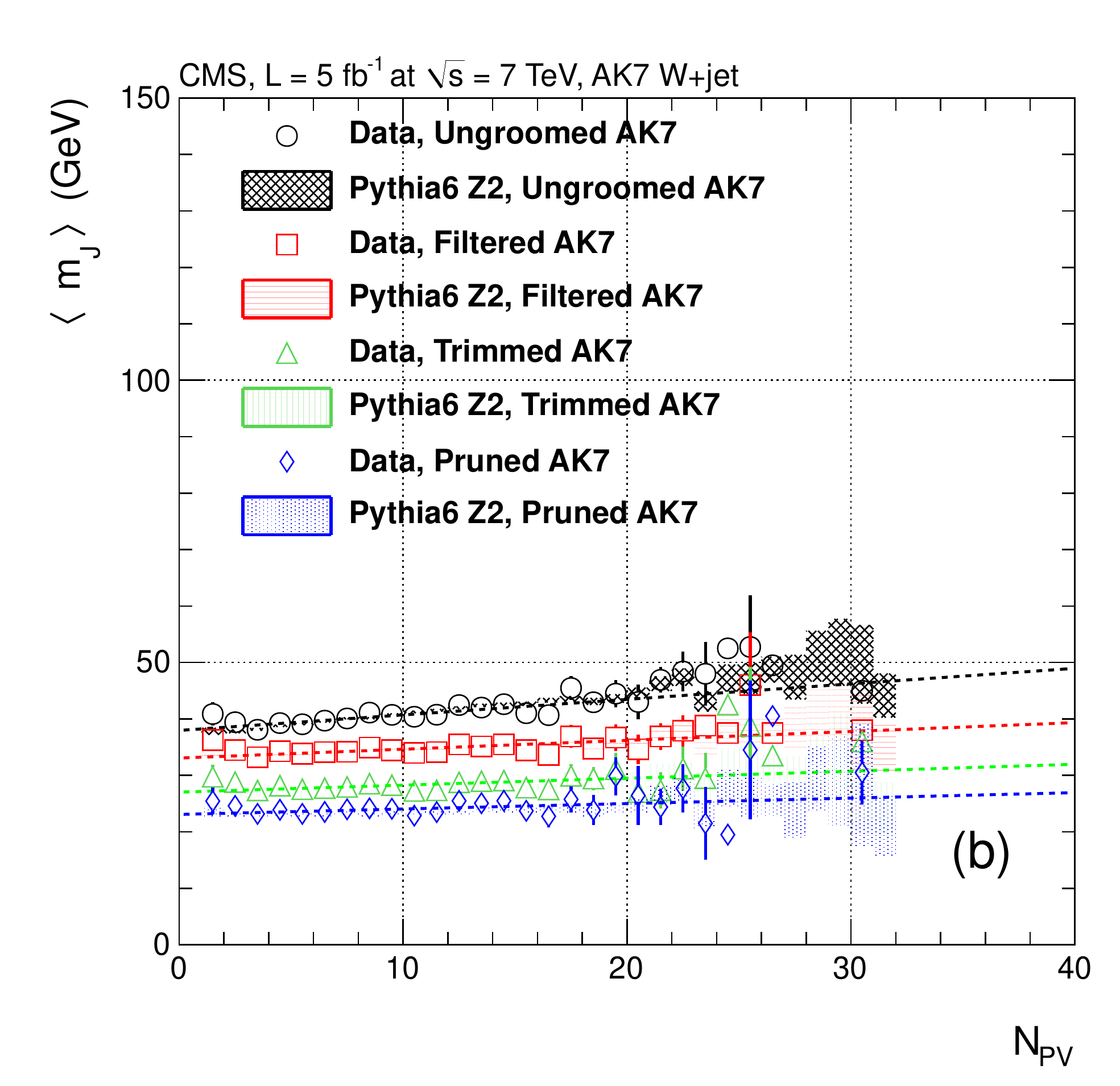}

\caption{Distributions of the average jet mass for AK jets as a function of the number of reconstructed primary vertices: (a) for different jet radii, and (b) for AK7 jets, comparing the impact of grooming algorithms to results without grooming.
\label{figs:histAK7MjetVsNvtx_nvtxPlots}}
\end{figure}

The mean jet mass $\langle m_J \rangle$ for AK jets is presented for
size parameters $R = 0.5$, 0.7, and 0.8, as a function of the total number of reconstructed
primary vertices ($N_{\mathrm{PV}}$) in
Fig.~\ref{figs:histAK7MjetVsNvtx_nvtxPlots}(a), for data and MC simulation.
The mean mass for $N_{\mathrm{PV}}=1$ increases
linearly with the jet radius from 0.5 to 0.8. A measure of the
dependence of $\langle m_J \rangle$ on pileup is given by the slope of a
linear fit to the jet mass versus $N_{\mathrm{PV}}$. The ratios of these
slopes ($s_R$) are found to be roughly consistent with the ratio of the third
power of the jet radius, as summarized in Table~\ref{tab:slopes}.

\begin{table}[!ht]
\topcaption{Slopes of linear fits of $\langle m_J \rangle$ as a function
  of $N_{\mathrm{PV}}$ for AK jets of different $R$ values.}
\label{tab:slopes}
\begin{center}
\begin{tabular}{ccc} \hline
Ratio of slopes & Measured & Expected \\
\hline\rule{0pt}{12pt}
$s_{0.7}/s_{0.5}$ & $2.7 \pm 0.9\stat$ & $(0.7/0.5)^3 = 2.74$  \\
$s_{0.8}/s_{0.5}$ & $3.3 \pm 1.0\stat$ & $(0.8/0.5)^3 = 4.10$  \\
$s_{0.8}/s_{0.7}$ & $1.2 \pm 0.2\stat$ & $(0.8/0.7)^3 = 1.49$  \\
\hline
\end{tabular}
\end{center}
\end{table}

\noindent This is in agreement with predictions for scaling of the mean mass~\cite{Dasgupta:2008}. The $R^3$ dependence can be understood in terms of the increase of the jet area as $R^2$. Simultaneously, the contribution of these particles to the jet mass scales with the distance between them, or ${\approx} R/2$, yielding another power of $R$.

In Fig.~\ref{figs:histAK7MjetVsNvtx_nvtxPlots}(b) we show the
dependence of $\langle m_J \rangle$ on $N_{\mathrm{PV}}$, for AK7 jets, for
different grooming algorithms. The grooming significantly reduces the
impact of pileup on $\langle m_J \rangle$, as reflected by the decrease
of the slope of the linear fit to the groomed-jet data points, as
summarized in Table~\ref{tab:slopes2}.

\begin{table}[!ht]
\topcaption{Values of slopes for the dependence of $\langle m_J \rangle$ on $N_{\mathrm{PV}}$ for AK jets with different radii and clustering algorithms.}
\label{tab:slopes2}
\begin{center}
\begin{tabular}{ccc} \hline
Jet R & Clustering algorithm & $s_R$ (\GeVns/PV) \\ \hline
AK5 & ungroomed & $0.10 \pm 0.03\stat$   \\
AK7 & ungroomed & $0.28 \pm 0.03\stat$  \\
AK7 & filtered & $0.16 \pm 0.02\stat$  \\
AK7 & trimmed & $0.12 \pm 0.04\stat$  \\
AK7 & pruned  & $0.10 \pm 0.05\stat$  \\
AK8 & ungroomed & $0.33 \pm 0.03\stat$  \\
\hline
\end{tabular}
\end{center}
\end{table}

The observed agreement between data and simulation in
Fig.~\ref{figs:histAK7MjetVsNvtx_nvtxPlots} provides
support for our characterization of jet grooming and pileup, and
the decrease in slopes suggests
that grooming is indeed an effective tool for suppressing the impact
of pileup on jets with large $R$ parameters.

\section{Corrections and systematic uncertainties}
\label{sec:systematics}

Before comparison of the jet mass distributions with QCD predictions, the data are
corrected to the particle level for detector effects, such as resolution and
acceptance. The simulated particle-level jets are
reconstructed with the same algorithm and with the same parameters as
the PF jets. We use the unfolding procedure described in
Refs.~\cite{unfolding_extra1,unfolding_extra2,unfolding_extra3,unfolding_extra4,agostini}
to correct the jet mass, through
an iterative technique for finding the
maximum-likelihood solution of the unfolding problem.
The detector response matrix is obtained in MC studies of jets.
In general, the number of iterations must be tuned to minimize the
impact of
statistical fluctuations on the result. In practice, however, the
procedure is largely insensitive to
the precise settings and binning of events and four iterations usually
suffice. A larger number of iterations were found to
provide the same results except for small fluctuations in the tails of
distributions. A simpler bin-by-bin unfolding is used as a
cross-check,
and is found to provide similar results, with fluctuations
in the tails of the distributions. The jet transverse momenta are
not unfolded.

Systematic uncertainties are estimated by modifying the
response matrix for each source of uncertainty by $\pm 1$ standard
deviation, and comparing the mass
distribution to the nominal results, based on simulated \PYTHIA
events. The difference in the unfolded mass spectrum from such a change is taken as
the uncertainty arising from that source.

The experimental uncertainties that can affect the unfolding
of the jet mass
include the jet energy scale (JES),
jet energy resolution (JER), and jet angular resolution (JAR).
The uncertainty from JES is estimated by raising and lowering the jet
four-momenta by the measured uncertainty as a function of jet $\pt$
and $\eta$~\cite{citeJEC}, which typically corresponds to 1--2\% for the jets
in this analysis. Two additional $\pt$- and $\eta$-independent
uncertainties are included: a 1\% uncertainty to account for
differences observed between the measured and
predicted $\wboson$ mass for high-$\pt$ jets in a $\ttbar$-enriched sample, and a
3\% uncertainty to account for differences in the groomed and
ungroomed energy responses found in MC simulation~\cite{EXO-11-006}.

The impact of uncertainties in JER and JAR on $m_J$ are evaluated by
smearing the jet energies, as well as the resolutions in $\eta$ and
$\phi$, each by 10\% in the MC simulation relative to the
particle-level generated jets~\cite{citeJEC}.
These estimated uncertainties on JER and JAR are found to be
essentially the same for all jet grooming techniques in MC studies.
Since this analysis uses jets constructed from PF constituents, the
charged particles have excellent energy and angular
resolutions, but their use induces a dependence on tracking
uncertainties, \eg, tracking efficiency. This dependence is accounted for
implicitly in the $\pm$10\% changes in jet energy and angular
resolutions, since such changes would lead to a difference between
expected and observed values of these quantities. The same is true for the
neutral electromagnetic
component of the jet (primarily from $\pi^0 \rightarrow \gamma\gamma$
decays).

The remaining sources of uncertainty are estimated from MC simulation.
The tracking information is not sensitive to the neutral hadronic
component of jets, and this small contribution is taken
directly from simulation.
We estimate this remaining uncertainty by comparing the unfolded data using \PYTHIA
and using \HERWIG, and assign the difference as a systematic uncertainty.
This also accounts for the uncertainty from modeling parton showers.
The latter effect often comprises the largest uncertainty in the unfolded jet mass
distributions as described below.
Other theoretical ambiguities that can affect the unfolding of the jet
mass include the variation of the parton distribution functions and
the modeling of initial and final-state radiation (ISR/FSR). The former
was investigated and found to be much smaller than the difference
between the unfolding with \PYTHIA and the unfolding with \HERWIG, and
hence is neglected. The latter is included implicitly in the uncertainty between
\PYTHIA and \HERWIG.

As described in Section~\ref{sec:reconstruction}, the jets used
in this analysis are reconstructed after removing
the charged hadrons that appear to emanate from subleading primary
vertices.
This procedure produces a dramatic (${\approx} 60\%$) reduction in the pileup
contribution to jets.
The residual uncertainty from pileup is obtained through MC simulation,
estimated by increasing and decreasing the cross section for minimum-bias events by 8\%.

In the dijet analysis, there can be incorrect assignments of leading
reconstructed jets relative to the generator level, \eg, two
generator-level jets can be matched to three reconstructed jets, or vice versa.
This effect causes a bias in the unfolding procedure, which becomes greater
at small $\pt$. This bias is corrected through MC studies
of matching of particle-level jets to reconstructed jets,
and the magnitude of the bias correction is also
added to the overall systematic uncertainty.
Such misassignments are negligible in the V+jet analysis.

\section{Results from dijet final states}

\label{sec:rawDataMCComparisons}
\label{sec:dijetresults}

\ifnpas
In this Section, the detector-level distributions of the jet mass
are investigated. Each of the distributions are made for
the $\pt^\mathrm{AVG}$ bins described in Section~\ref{sec:ptBinAssignment}.
Each of the figures shows the raw event counts per $\pt^\mathrm{AVG}$ bin,
as a function of the average jet mass $m_J^\mathrm{AVG}$.

Figure~\ref{figs:histAK7PtAvgVsMjetGroomOverReco_ratioPlots}
shows a comparison of the jet mass from the groomed jets
divided by the jet mass of matched ungroomed jets, for the
three grooming techniques, for both data and the \PYTHIA Monte Carlo.
The data and the MC both exhibit similar behavior. In general,
the filtering algorithm (black) is the least aggressive grooming technique,
with groomed jet masses close to the ungroomed case.
The trimming algorithm (red) is moderately aggressive, and the
pruning algorithm (blue) is the most aggressive. In the case of
the pruning algorithm, a bimodal distribution begins to manifest,
which is typical of this algorithm since the parameters we have
chosen require two subjets to be created. In the cases where
the pruned jet mass is close to the ungroomed jet mass,
jets usually have large ``core'' components
and small amounts of radiation, whereas when the pruned jet
mass is closer to 0, the jets are more symmetrically split
due to gluons splitting into two jets that fall within
our $D=0.7$ parameter.
\fi

\ifnpas
Figures~\ref{figs:histAK7MjetVsPtAvg_rawDataMCComparisons_stacktrigs_pt_2}-
\ref{figs:histAK7MjetVsPtAvg_rawDataMCComparisons_stacktrigs_pt_5}
show the jet mass distribution for AK7 jets, along with
the trigger breakdown for all of the $\pt^\mathrm{AVG}$ bins. In each case,
only one trigger contributes to any given $\pt^\mathrm{AVG}$ bin.

Figures~\ref{figs:histAK7MjetVsPtAvg_rawDataMCComparisons_pt_2}-
\ref{figs:histAK7MjetVsPtAvg_rawDataMCComparisons_pt_2_Pruned}
show the detector-level distributions for the various jet grooming
algorithms, compared to predictions from \PYTHIA, \PYTHIA8, and
\HERWIG Monte Carlo samples. Each MC sample is normalized
according to the expected luminosity and computed cross sections.
\fi

\ifnpas

\begin{figure}[htbp]
\centering
\includegraphics[width=0.95\textwidth]{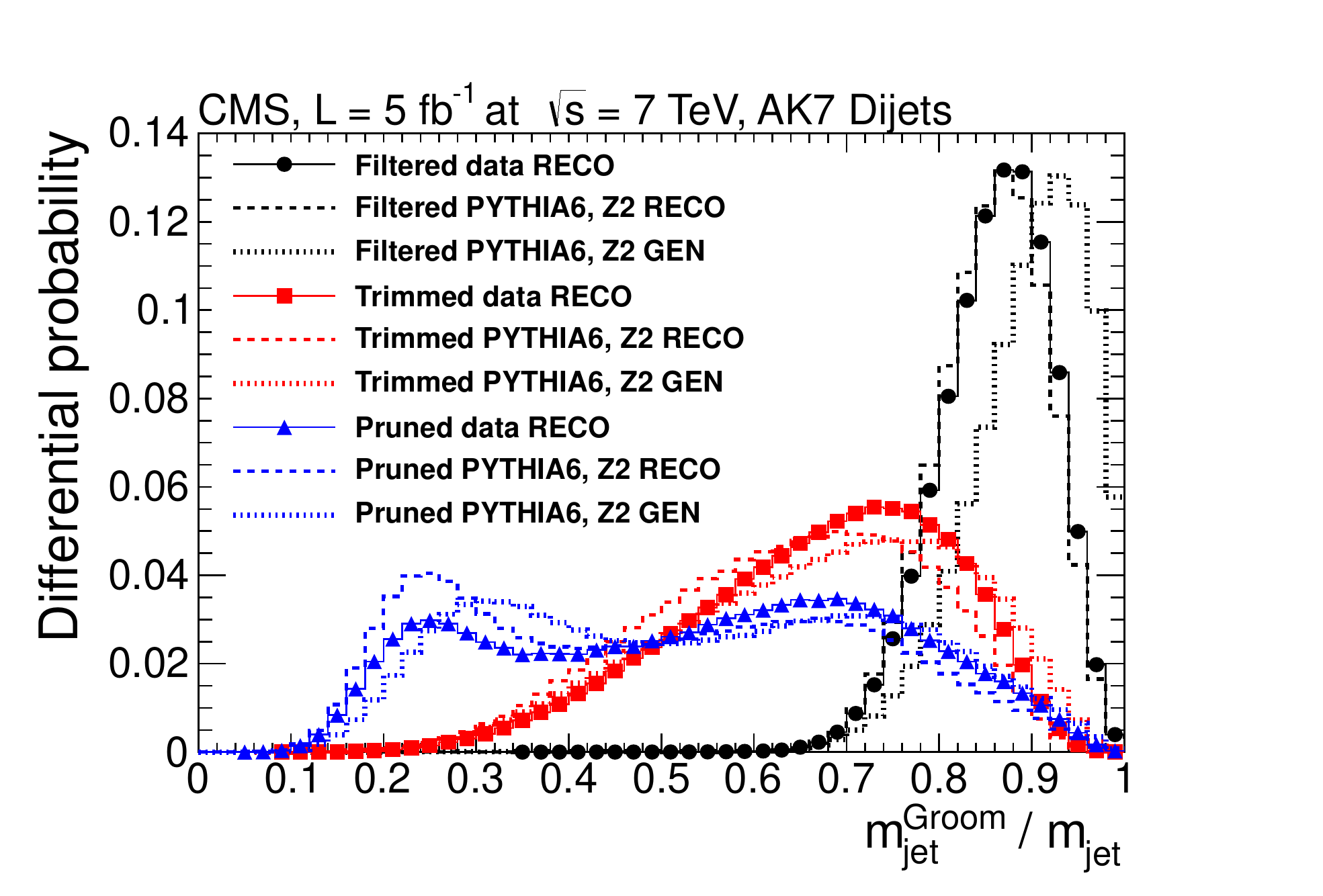}
\caption{Comparison of the jet mass from the groomed jets
divided by the jet mass of matched ungroomed jets for the
three grooming techniques, for both data and the \PYTHIA Monte Carlo.
\label{figs:histAK7PtAvgVsMjetGroomOverReco_ratioPlots}}
\end{figure}

\fi

\label{sec:results}

\ifnpas
The unfolded distributions of the
averse jet mass are also investigated.
In Figures~\ref{figs:unfoldedMeasurementDijets_1_allsys}-
\ref{figs:unfoldedMeasurementDijets_9_Pruned_allsys},
the differential cross
section from Eq.(\ref{eq:pdf_mjet_simple}) is shown for
various $\pt^\mathrm{AVG}$ bins, as a function of
$m_{J}^\mathrm{AVG}$, as described in
Section~\ref{sec:dataSampleAndEventSelection},
for ungroomed jets and for different grooming algorithms.
As described in Eq.(\ref{eq:pdf_mjet_simple}),
each distribution in each $\pt^\mathrm{AVG}$ is separately normalized to
unity, and each bin content is divided by the bin width of the
$m_J^\mathrm{AVG}$ bin,
so these plots show the probability distribution function of $m_J^\mathrm{AVG}$,
with units of 1/\GeVns (see Eq.(~\ref{eq:pdf_mjet_simple})).
The shapes for $m_J^\mathrm{AVG}$ in the MC samples are
taken directly from the MC.

The statistical uncertainty is shown in light shading, the uncertainties due to the jet-energy resolution, jet-energy scale, and jet-angular resolution are shown in shades of brown, the uncertainty due to pileup is shown in green, and the uncertainty due to the parton shower differences are shown in dark shading.
The simulated distribution from \PYTHIA is shown in solid lines,
from \PYTHIAEIGHT in dashed lines, and from \HERWIG in dotted lines.
The bottom frame shows the ratio of the true distribution from
the simulation divided by the unfolded distribution, along with
the uncertainties in the unfolded distribution.

Figures~\ref{figs:unfoldedMeasurementDijets_1}-
\ref{figs:unfoldedMeasurementDijets_9_Pruned},
show the same plots, however in a simplified format with only the
total and statistical uncertainties shown in dark shading and
light shading, respectively.
\fi

\begin{figure}[htbp]
\centering
\includegraphics[width=0.95\textwidth]{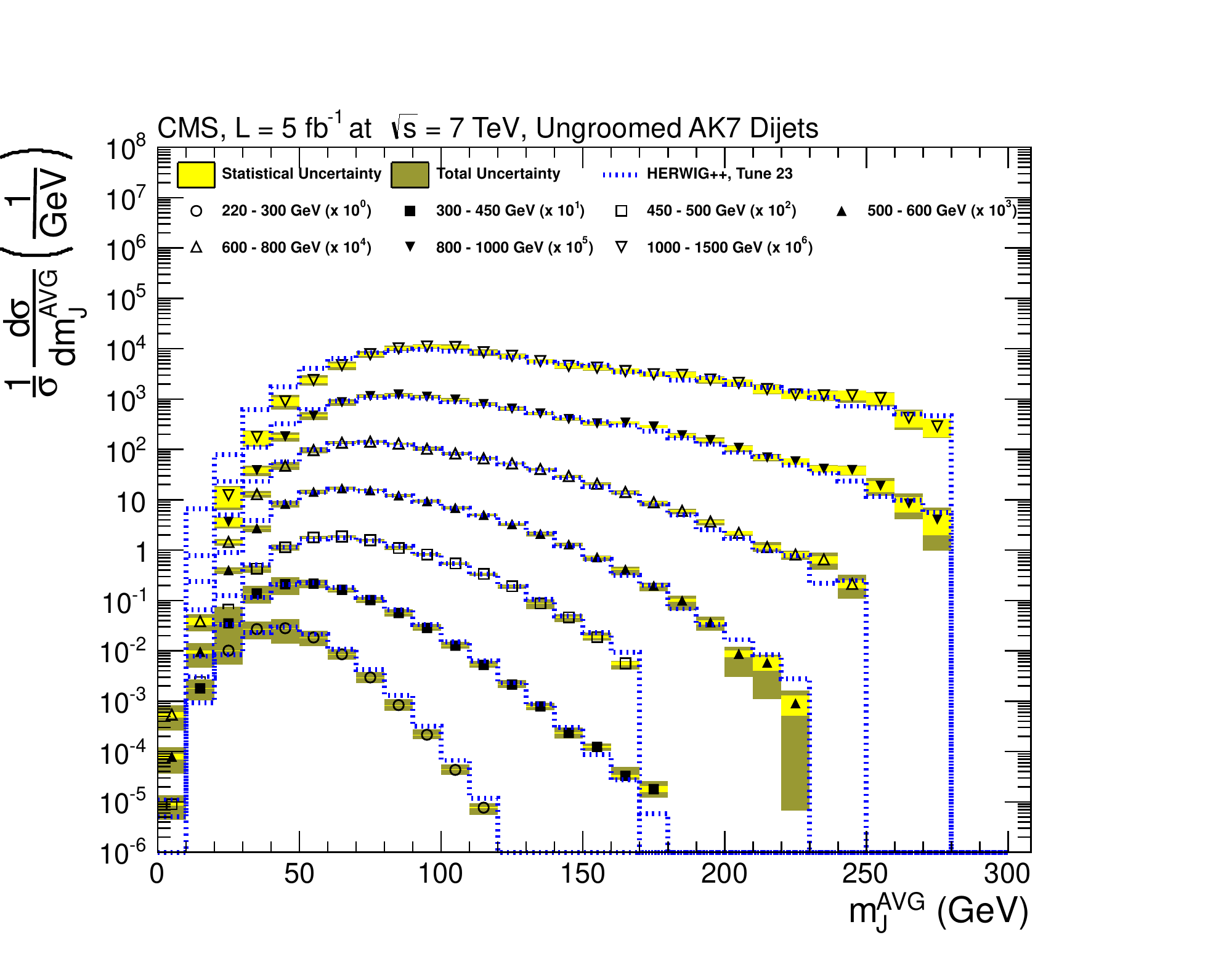}
\caption{Unfolded distributions for the mean mass of the two leading jets in dijet events for reconstructed AK7 jets,
separated according to intervals in $\pt^\mathrm{AVG}$ (the mean $\pt$ of the two jets).
The data are shown by the symbols indicating different bins in the mean $\pt$ of the two jets.
The statistical uncertainty is shown in light shading, and the
total uncertainty in dark shading.
Predictions
from \HERWIG are given by the dotted lines.
To enhance visibility, the distributions for larger values of $\pt^\mathrm{AVG}$
are scaled up by the factors given in the legend.
\label{figs:unfoldedMeasurementDijets_all}}
\end{figure}

\begin{figure}[htbp]
\centering
\includegraphics[width=0.95\textwidth]{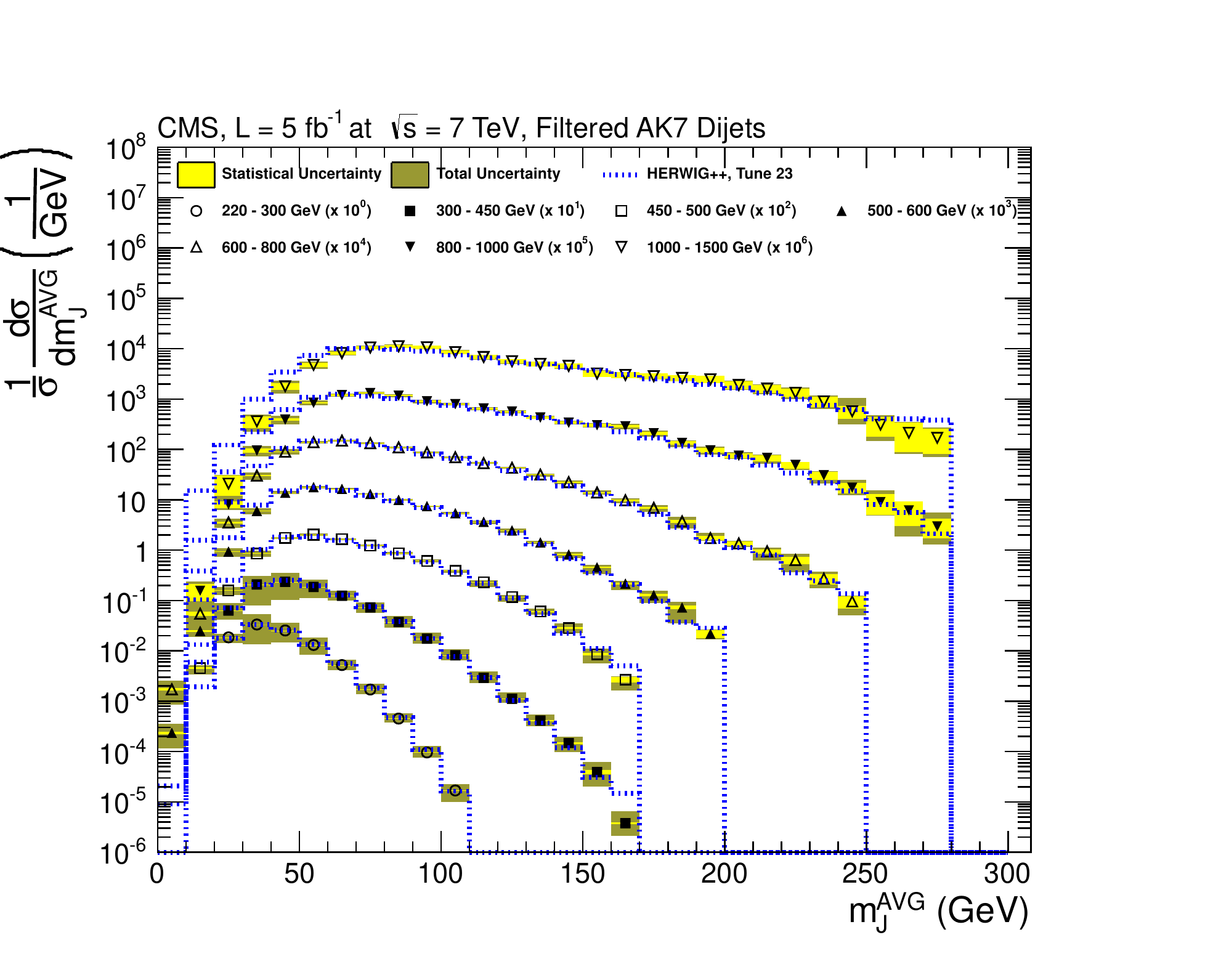}
\caption{Unfolded distributions for the mean mass of the two leading jets in dijet events for reconstructed filtered AK7 jets,
separated according to intervals in $\pt^\mathrm{AVG}$ (the mean $\pt$ of the two jets).
The data are shown by the symbols indicating different bins in the mean $\pt$ of the two jets.
The statistical uncertainty is shown in light shading, and the
total uncertainty in dark shading.
Predictions
from \HERWIG are given by the dotted lines.
To enhance visibility, the distributions for larger values of $\pt^\mathrm{AVG}$
are scaled up by the factors given in the legend.
\label{figs:unfoldedMeasurementDijets_all_Filtered}}
\end{figure}

\begin{figure}[htbp]
\centering
\includegraphics[width=0.95\textwidth]{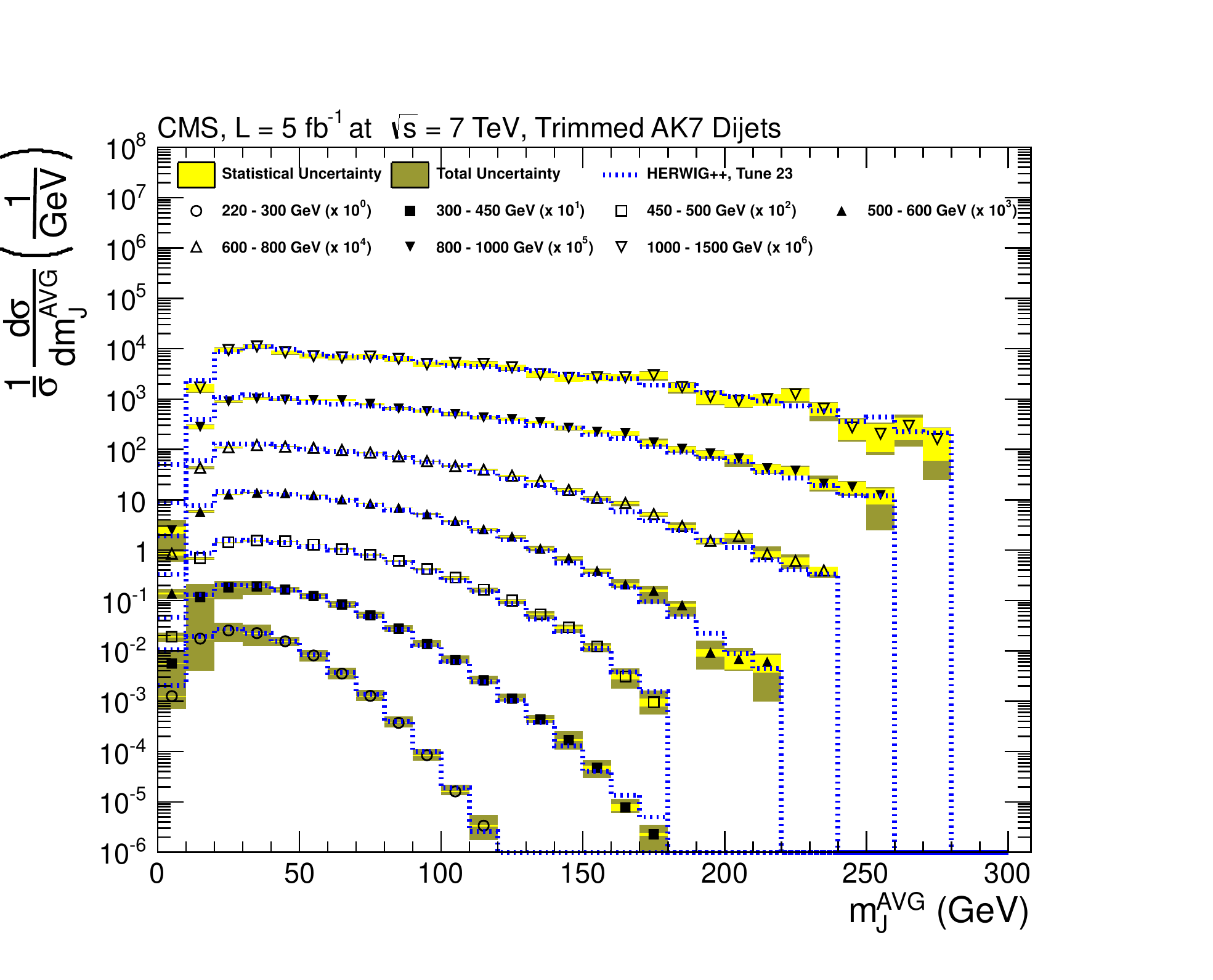}
\caption{Unfolded distributions for the mean mass of the two leading jets in dijet events for reconstructed trimmed AK7 jets,
separated according to intervals in $\pt^\mathrm{AVG}$ (the mean $\pt$ of the two jets).
The data are shown by the symbols indicating different bins in the mean $\pt$ of the two jets.
The statistical uncertainty is shown in light shading, and the
total uncertainty in dark shading.
Predictions
from \HERWIG are given by the dotted lines.
To enhance visibility, the distributions for larger values of $\pt^\mathrm{AVG}$
are scaled up by the factors given in the legend.
\label{figs:unfoldedMeasurementDijets_all_Trimmed}}
\end{figure}

\begin{figure}[htbp]
\centering
\includegraphics[width=0.95\textwidth]{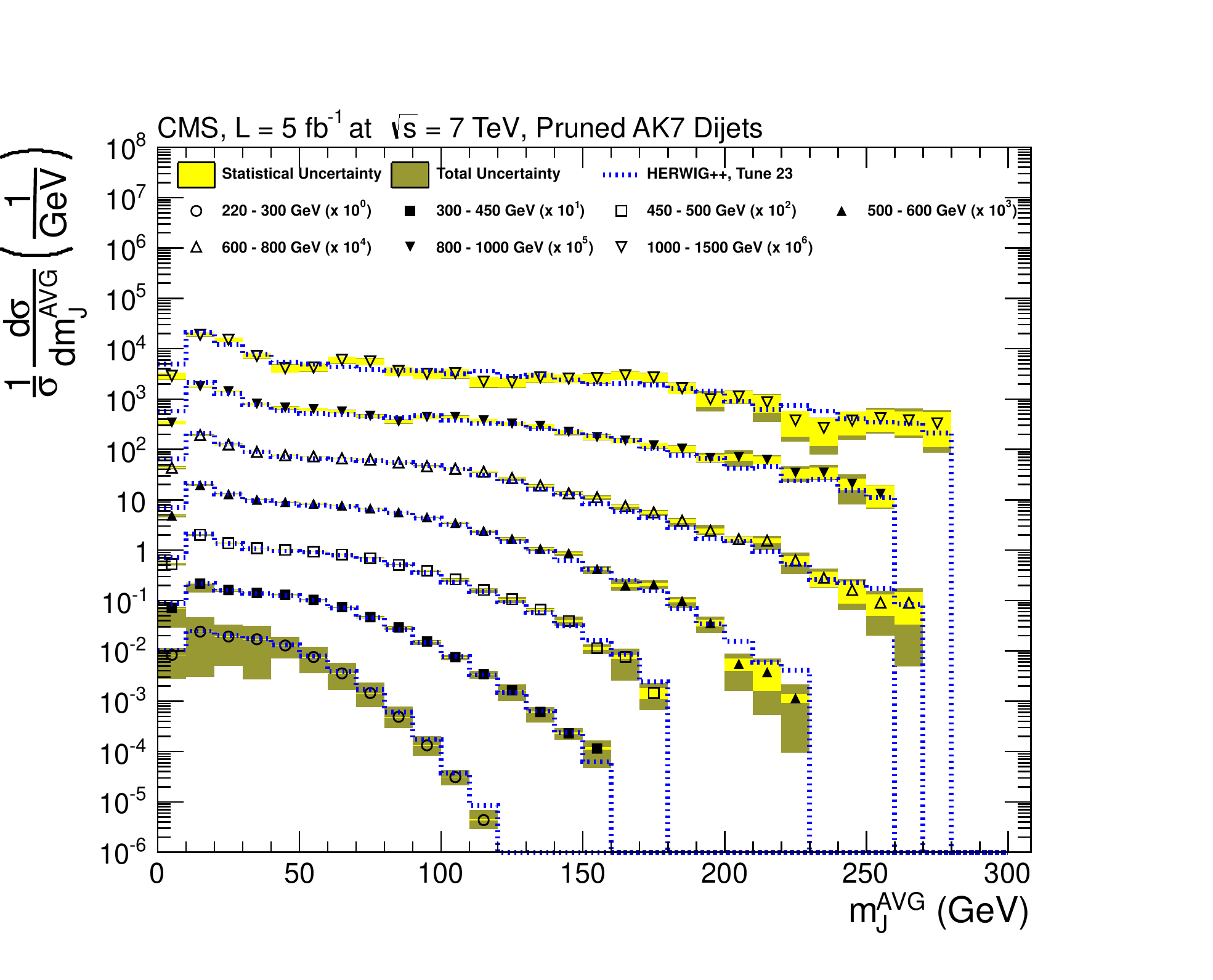}
\caption{Unfolded distributions for the mean mass of the two leading jets in dijet events for reconstructed pruned AK7 jets,
separated according to intervals in $\pt^\mathrm{AVG}$ (the mean $\pt$ of the two jets).
The data are shown by the symbols indicating different bins in the mean $\pt$ of the two jets.
The statistical uncertainty is shown in light shading, and the
total uncertainty in dark shading.
Predictions
from \HERWIG are given by the dotted lines.
To enhance visibility, the distributions for larger values of $\pt^\mathrm{AVG}$
are scaled up by the factors given in the legend.
\label{figs:unfoldedMeasurementDijets_all_Pruned}}
\end{figure}

\begin{figure}[htbp]
\centering
\includegraphics[width=0.95\textwidth]{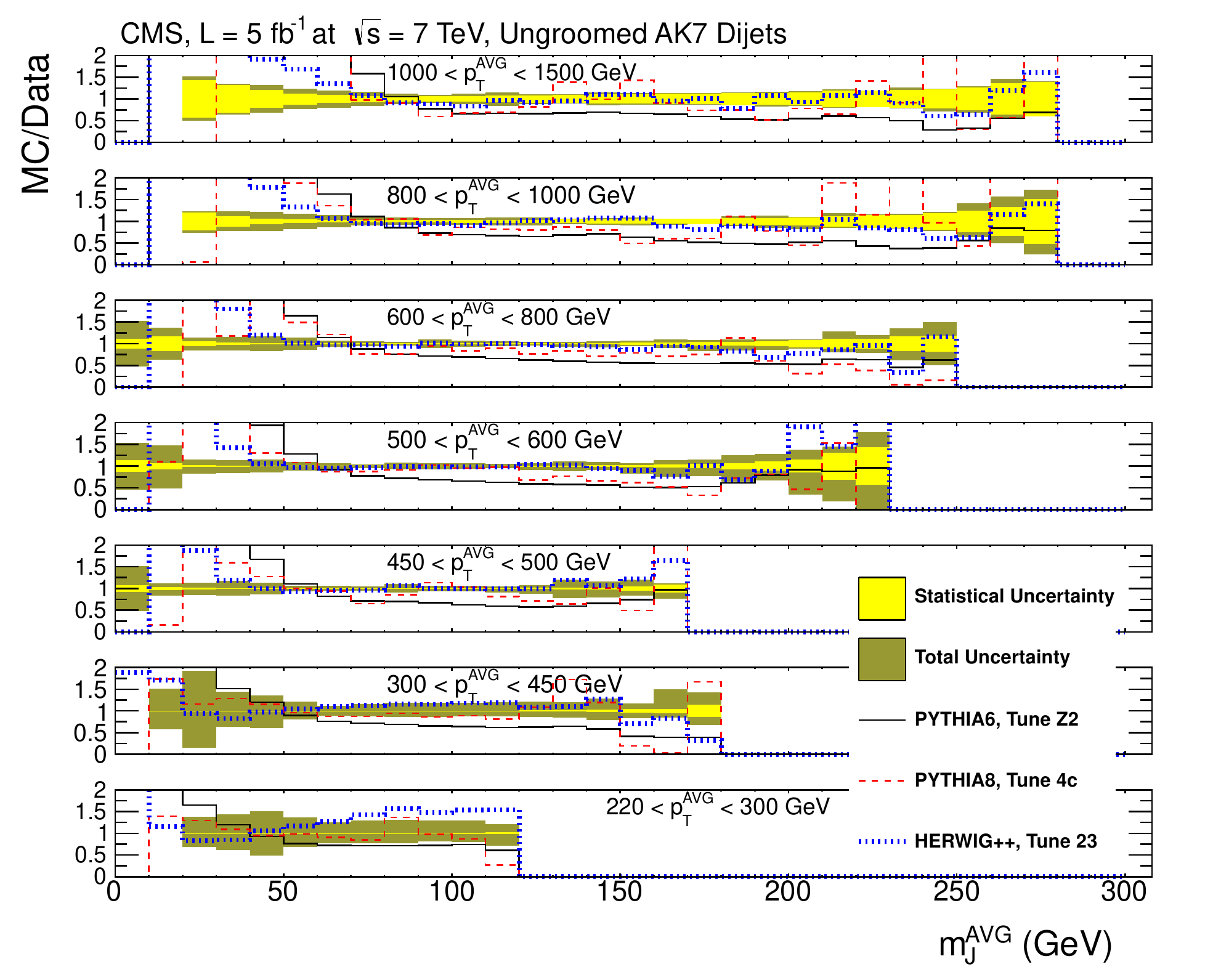}
\caption{Ratio of MC simulation to unfolded distributions of the jet mass for AK7 jets for the seven bins in $\pt^\mathrm{AVG}$.
The statistical uncertainty is shown in light shading, and the
total uncertainty is shown in dark shading.
The comparison for \PYTHIA is shown in solid lines, for \PYTHIAEIGHT in dashed lines, and for \HERWIG in dotted lines.
\label{figs:unfoldedMeasurementDijets_allfrac}}
\end{figure}

\begin{figure}[htbp]
\centering
\includegraphics[width=0.95\textwidth]{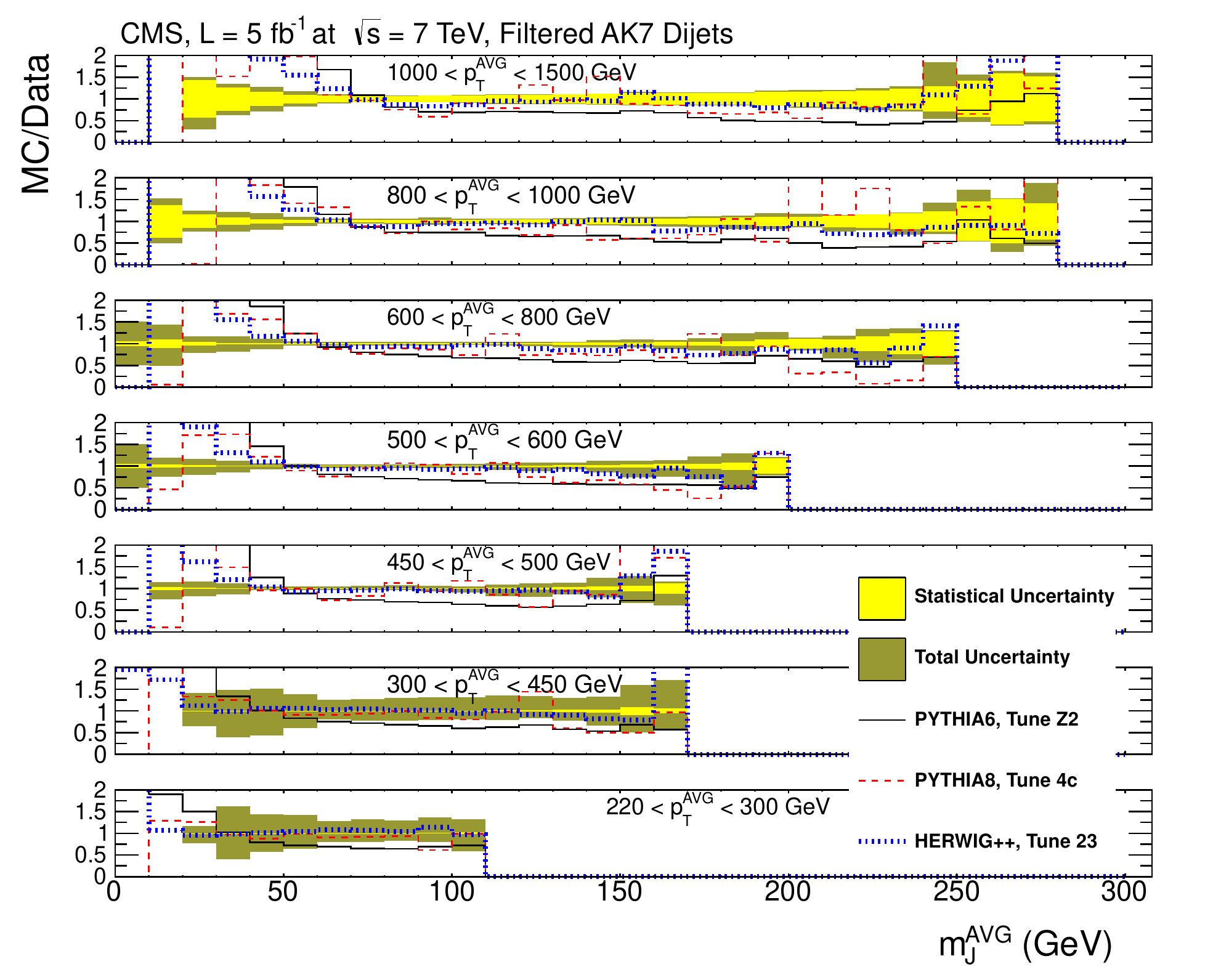}
\caption{Ratio of MC simulation to unfolded distributions of the jet mass for filtered AK7 jets for the seven bins in $\pt^\mathrm{AVG}$.
The statistical uncertainty is shown in light shading, and the
total uncertainty is shown in dark shading.
The comparison for \PYTHIA is shown in solid lines, for \PYTHIAEIGHT in dashed lines, and for \HERWIG in dotted lines.
\label{figs:unfoldedMeasurementDijets_allfrac_Filtered}}
\end{figure}

\begin{figure}[htbp]
\centering
\includegraphics[width=0.95\textwidth]{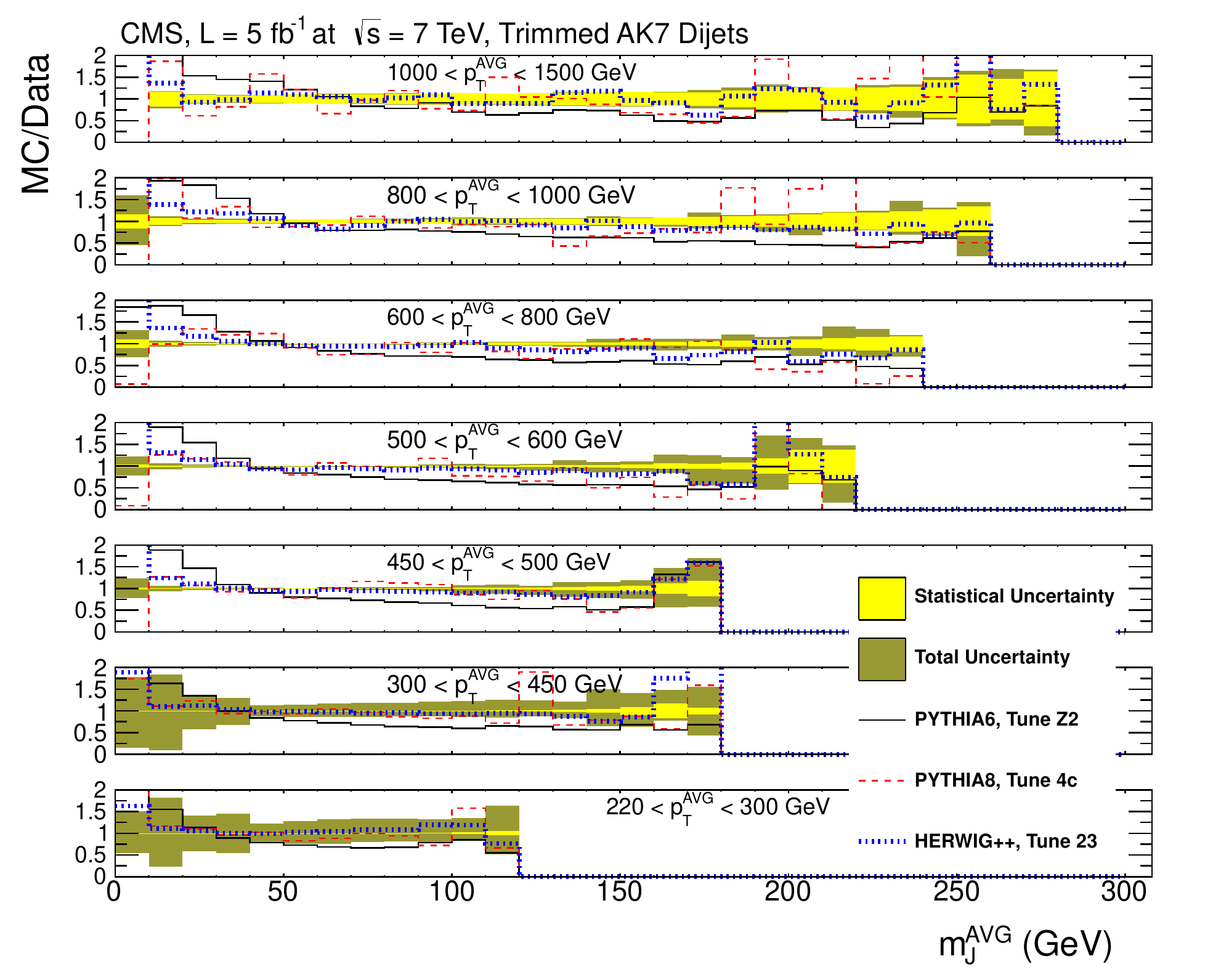}
\caption{Ratio of MC simulation to unfolded distributions of the jet mass for trimmed AK7 jets for the seven bins in $\pt^\mathrm{AVG}$.
The statistical uncertainty is shown in light shading, and the
total uncertainty is shown in dark shading.
The comparison for \PYTHIA is shown in solid lines, for \PYTHIAEIGHT in dashed lines, and for \HERWIG in dotted lines.
\label{figs:unfoldedMeasurementDijets_allfrac_Trimmed}}
\end{figure}

\begin{figure}[htbp]
\centering
\includegraphics[width=0.95\textwidth]{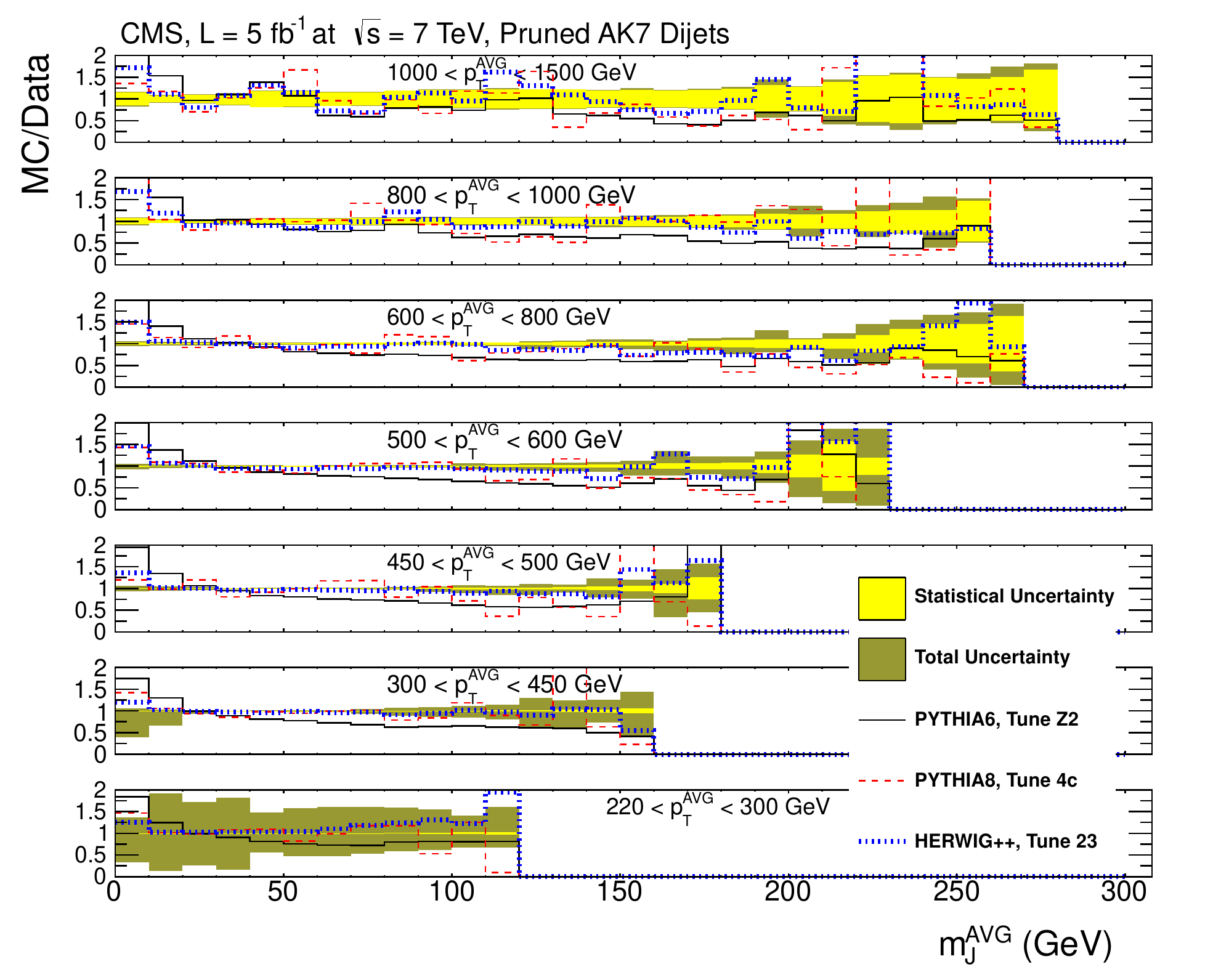}
\caption{Ratio of MC simulation to unfolded distributions of the jet mass for pruned AK7 jets for the seven bins in $\pt^\mathrm{AVG}$.
The statistical uncertainty is shown in light shading, and the
total uncertainty is shown in dark shading.
The comparison for \PYTHIA is shown in solid lines, for \PYTHIAEIGHT in dashed lines, and for \HERWIG in dotted lines.
\label{figs:unfoldedMeasurementDijets_allfrac_Pruned}}
\end{figure}

\ifpas

The differential probability distributions of Eq.~(\ref{eq:pdf_mjet_simple})
for $m_J^\mathrm{AVG}$ of the two leading jets in dijet events,
corrected for detector effects in the jet mass, are displayed in
Figs.~\ref{figs:unfoldedMeasurementDijets_all}--\ref{figs:unfoldedMeasurementDijets_all_Pruned}
for seven bins in $\pt^\mathrm{AVG}$ along with the \HERWIG predictions..
The $\pt^\mathrm{AVG}$ is
not corrected to the particle level, because the correction is
expected to be negligible for the momenta considered.
Results are shown
for ungroomed jets and for the three categories of grooming.
Each distribution is normalized to unity.
The ratios of the MC simulations used in
Figs.~\ref{figs:unfoldedMeasurementDijets_all}--\ref{figs:unfoldedMeasurementDijets_all_Pruned}
to the results for data, for \PYTHIA, \PYTHIAEIGHT, and for \HERWIG are given in Figs.~\ref{figs:unfoldedMeasurementDijets_allfrac}--\ref{figs:unfoldedMeasurementDijets_allfrac_Pruned}, respectively.

\fi

The largest systematic uncertainty is from the choice of parton-shower modeling used to calculate detector corrections, with
small, but still significant uncertainties arising from jet energy scale and resolution, and
small contributions from jet angular resolution and pileup.
In the 220--300\GeV and 300--450\GeV jet-$\pt$ bins, the $m_J < 50\GeV$ region
is dominated by uncertainties from unfolding (50--100\%), which are
negligible for $\pt^\mathrm{AVG}>450\GeV$.
For $m_J > 50\GeV$, the JES, JER, JAR, and pileup uncertainties each
contribute ${\approx} 10\%$. For the
450--1000\GeV $\pt$ bins, parton
showering dominates the uncertainties, which is around 50--100\% below
the peak of the $m_J$
distribution and 5--10\% for the rest of the distribution. For
$\pt > 1000\GeV$, statistical uncertainty dominates the entire mass range.

For clarity, the distributions in
Figs.~\ref{figs:unfoldedMeasurementDijets_allfrac}--\ref{figs:unfoldedMeasurementDijets_allfrac_Pruned}
are truncated where few events are recorded.
Bins in $m_J^\mathrm{AVG}$ with uncertainties of $>$ 100\% are
ignored to avoid overlap with more precise measurements in other
$\pt^\mathrm{AVG}$ bins.
The agreement with \HERWIG modeling of parton showers appears to be
best for $\pt^\mathrm{AVG}>300\GeV$ and $m_{J}^\mathrm{AVG}>20\GeV$.
However, the ungroomed and filtered jets show worse agreement
for $20<m_{J}^\mathrm{AVG}<50\GeV$ when $\pt^\mathrm{AVG}>450\GeV$.
For all generators and all $\pt^\mathrm{AVG}$ bins, the agreement is better
at larger jet masses. The disagreement is largest at the very lowest
mass values, which correspond to the region most sensitive to
the underlying event description and pileup,
and where the amount of showering is apparently underestimated in the
simulation.

\section{Results from V+jet final states}

\label{sec:vjetresults}

This section provides the probability density distributions as functions of
the mass of the leading jet in V+jet events. These distributions are
corrected for detector effects in the jet mass, and are compared to MC expectations
from \MADGRAPH (interfaced to \PYTHIA) and \HERWIG.
The jet mass distributions are studied in different ranges of $\pt$
between $125$ and $450\GeV$, as given in Table~\ref{tab:ptBins}.
(Just as in
the dijet results, $\pt$
is not corrected to the particle level.) For
jets reconstructed with the CA algorithm ($R=1.2$), we study only the
events with
$\pt>150\GeV$, which is most interesting for
heavy particle searches in the highly-boosted regime, where all decay
products are contained within $R=1.2$ jets~\cite{boostedHiggs}.

For clarity, the distributions are also truncated at large mass values where few events are recorded.
 Jet-mass bins with relative uncertainties
$>$ 100\% are also ignored to minimize overlap with more precise measurements in other $\pt$ bins.

\begin{figure}[!htb]
\includegraphics[width=0.99\textwidth]{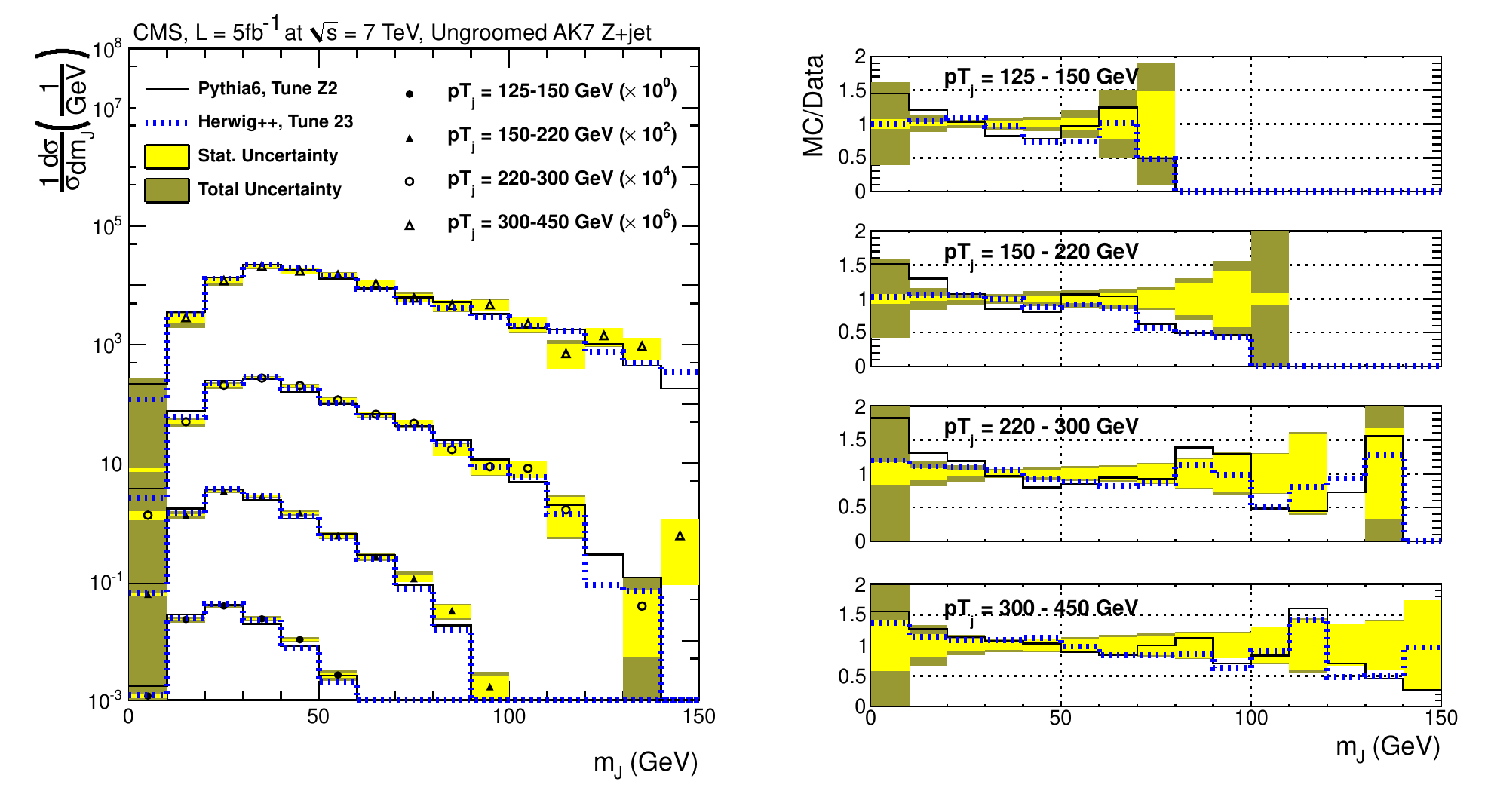}
\caption{Unfolded, ungroomed AK7 $m_J$ distribution for $\Z(\ell\ell)$+jet events. The data (black symbols) are compared to MC expectations from {\MADGRAPH}+\PYTHIA (solid lines) and \HERWIG (dotted lines) on the left. The ratio of MC to data is given on the right.
The statistical uncertainty is shown in light shading, and the total uncertainty in dark shading.}
\label{figs:AK7ZmmInt1}
\end{figure}

\begin{figure}[!htb]
\includegraphics[width=0.99\textwidth]{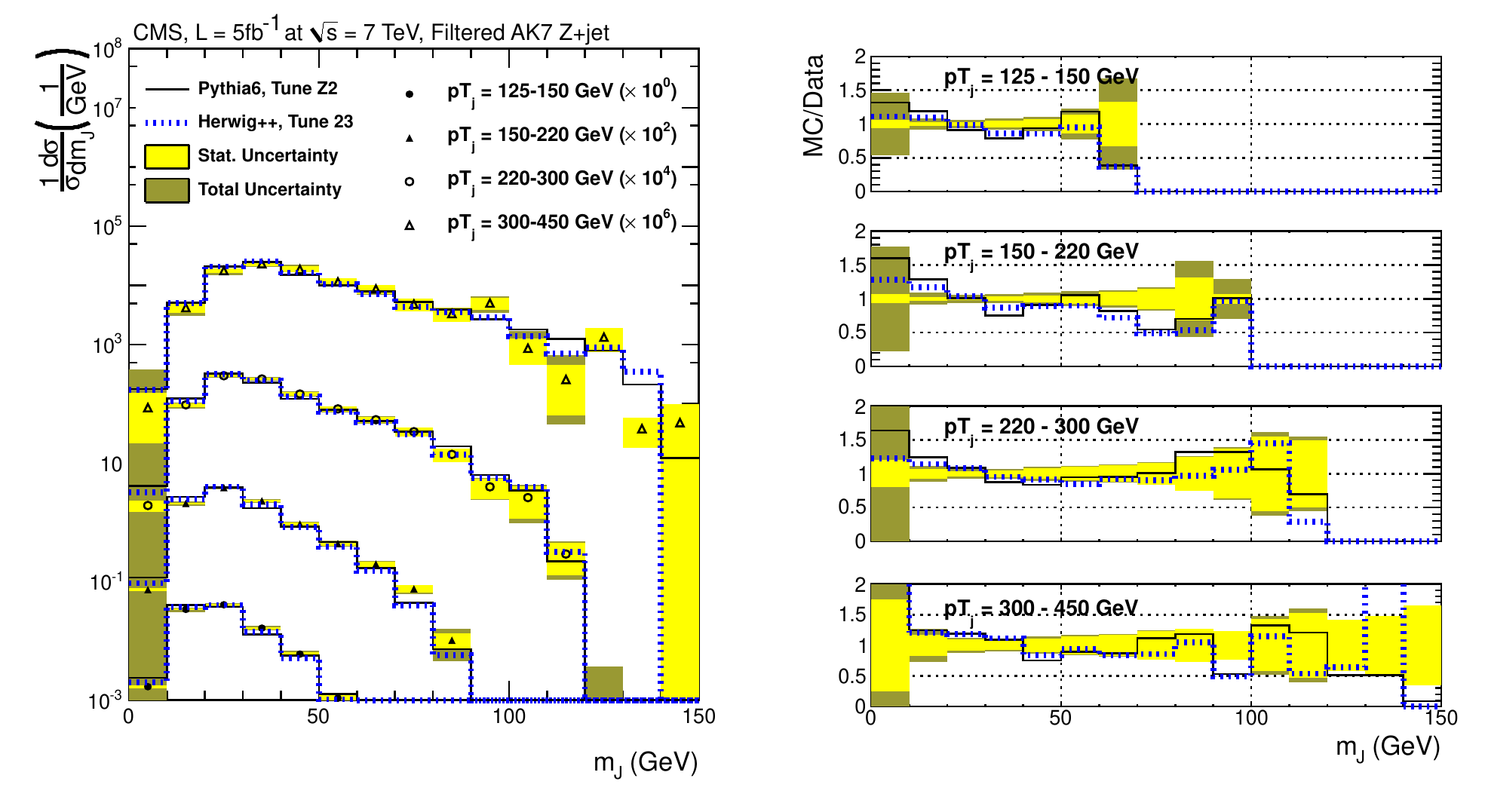}
\caption{Unfolded AK7 filtered $m_J$ distribution for $\Z(\ell\ell)$+jet events. The data (black symbols) are compared to MC expectations from {\MADGRAPH}+\PYTHIA (solid lines) and \HERWIG (dotted lines) on the left. The ratio of MC to data is given on the right.
The statistical uncertainty is shown in light shading, and the total uncertainty in dark shading.}
\label{figs:AK7ZmmInt3}
\end{figure}

\begin{figure}[!htb]
\includegraphics[width=0.99\textwidth]{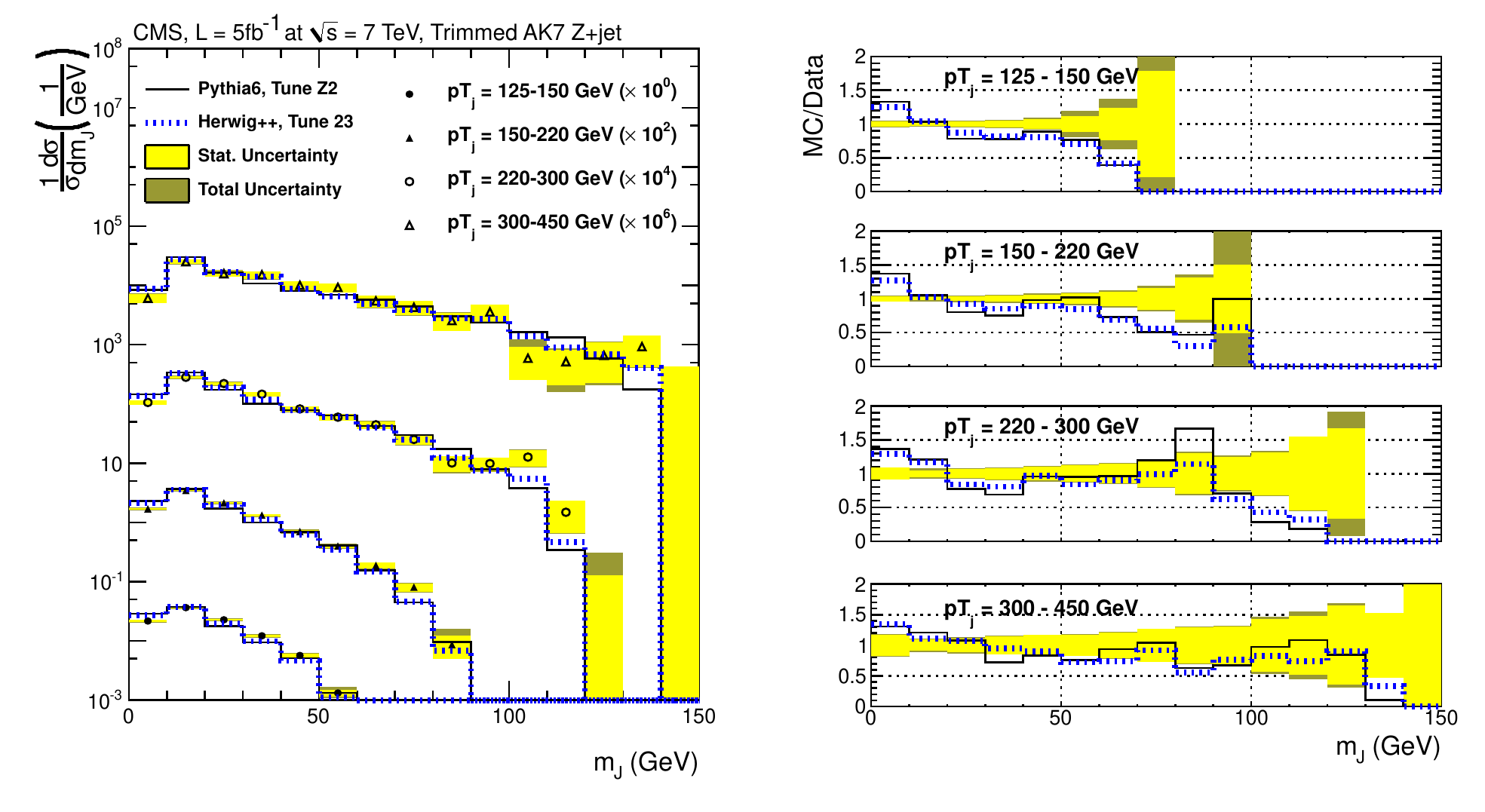}
\caption{Unfolded AK7 trimmed $m_J$ distribution for $\Z(\ell\ell)$+jet events. The data (black symbols) are compared to MC expectations from {\MADGRAPH}+\PYTHIA (solid lines) and \HERWIG (dotted lines) on the left. The ratio of MC to data is given on the right.
The statistical uncertainty is shown in light shading, and the total uncertainty in dark shading.}
\label{figs:AK7ZmmInt4}
\end{figure}

\begin{figure}[!htb]
\includegraphics[width=0.99\textwidth]{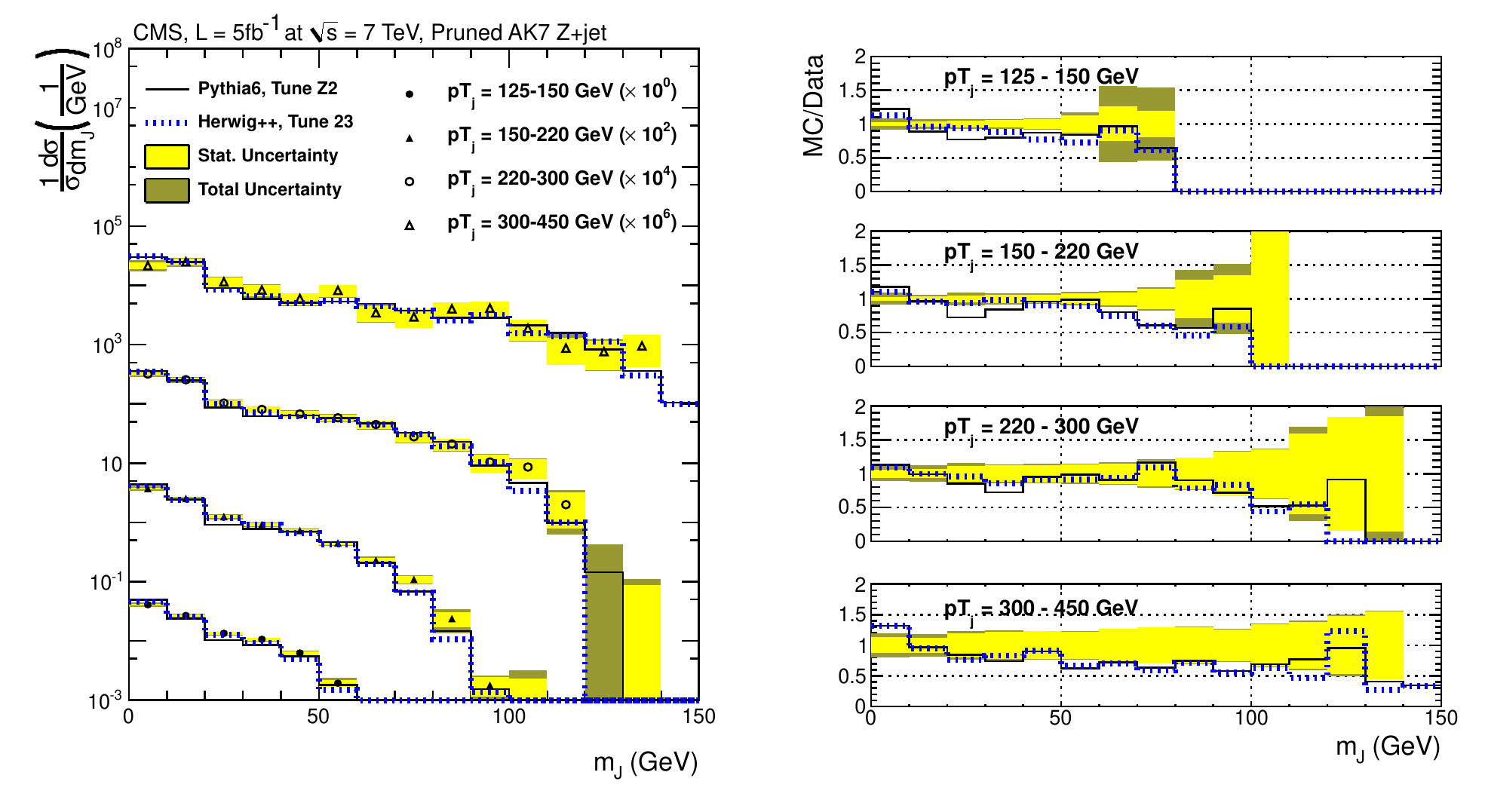}
\caption{Unfolded AK7 pruned $m_J$ distribution for $\Z(\ell\ell)$+jet events. The data (black symbols) are compared to MC expectations from {\MADGRAPH}+\PYTHIA (solid lines) and \HERWIG (dotted lines) on the left. The ratio of MC to data is given on the right.
The statistical uncertainty is shown in light shading, and the total uncertainty in dark shading.}
\label{figs:AK7ZmmInt2}
\end{figure}

\begin{figure}[!htb]
\centering
\includegraphics[width=0.99\textwidth]{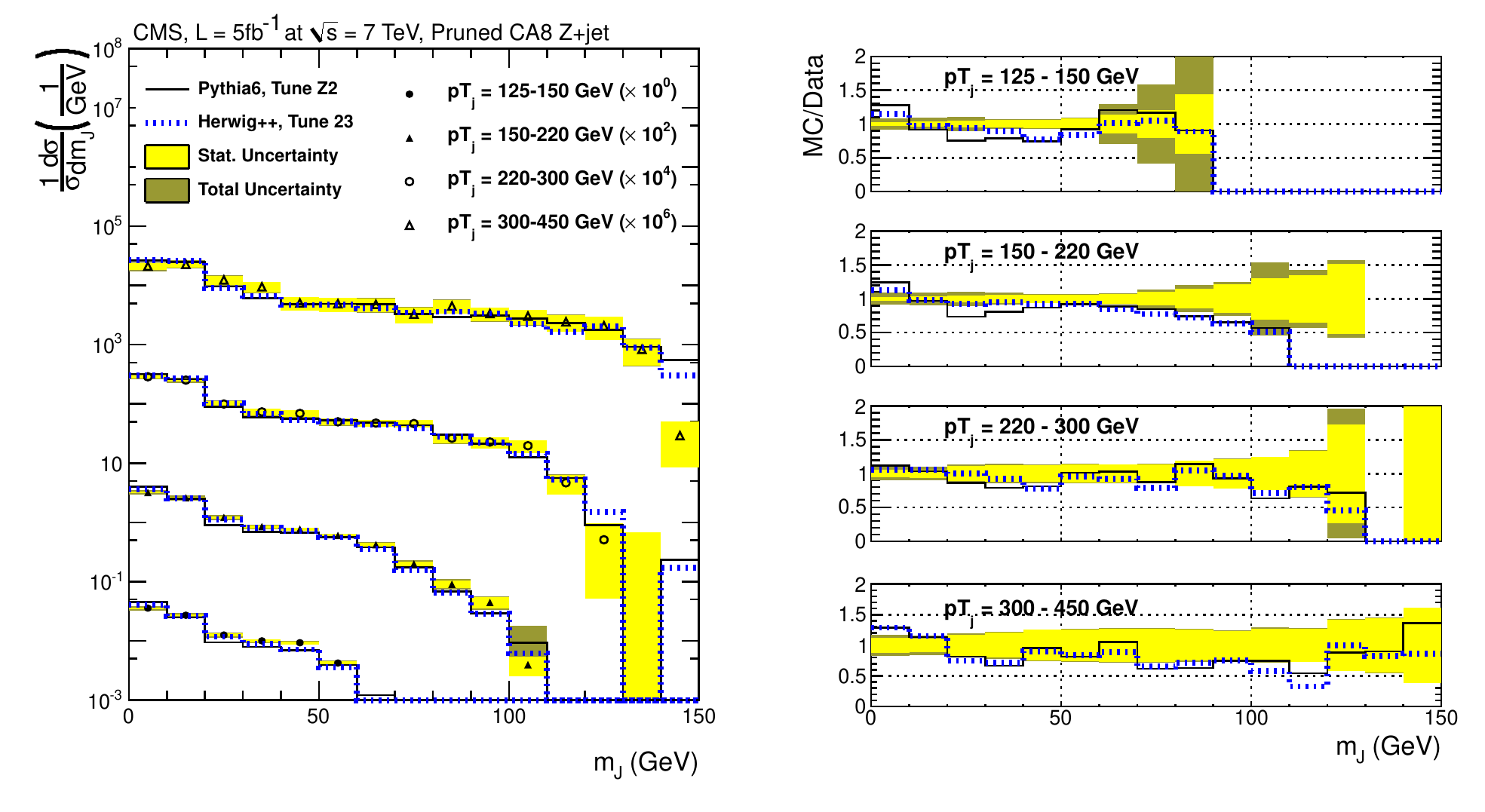}
\caption{Unfolded CA8 pruned $m_J$ distribution for $\Z(\ell\ell)$+jet events. The data (black symbols) are compared to MC expectations from {\MADGRAPH}+\PYTHIA (solid lines) and \HERWIG (dotted lines) on the left. The ratio of MC to data is given on the right.
The statistical uncertainty is shown in light shading, and the total uncertainty in dark shading.}
\label{figs:prunedZmmInt1}
\end{figure}

\begin{figure}[!htb]
\centering
\includegraphics[width=0.99\textwidth]{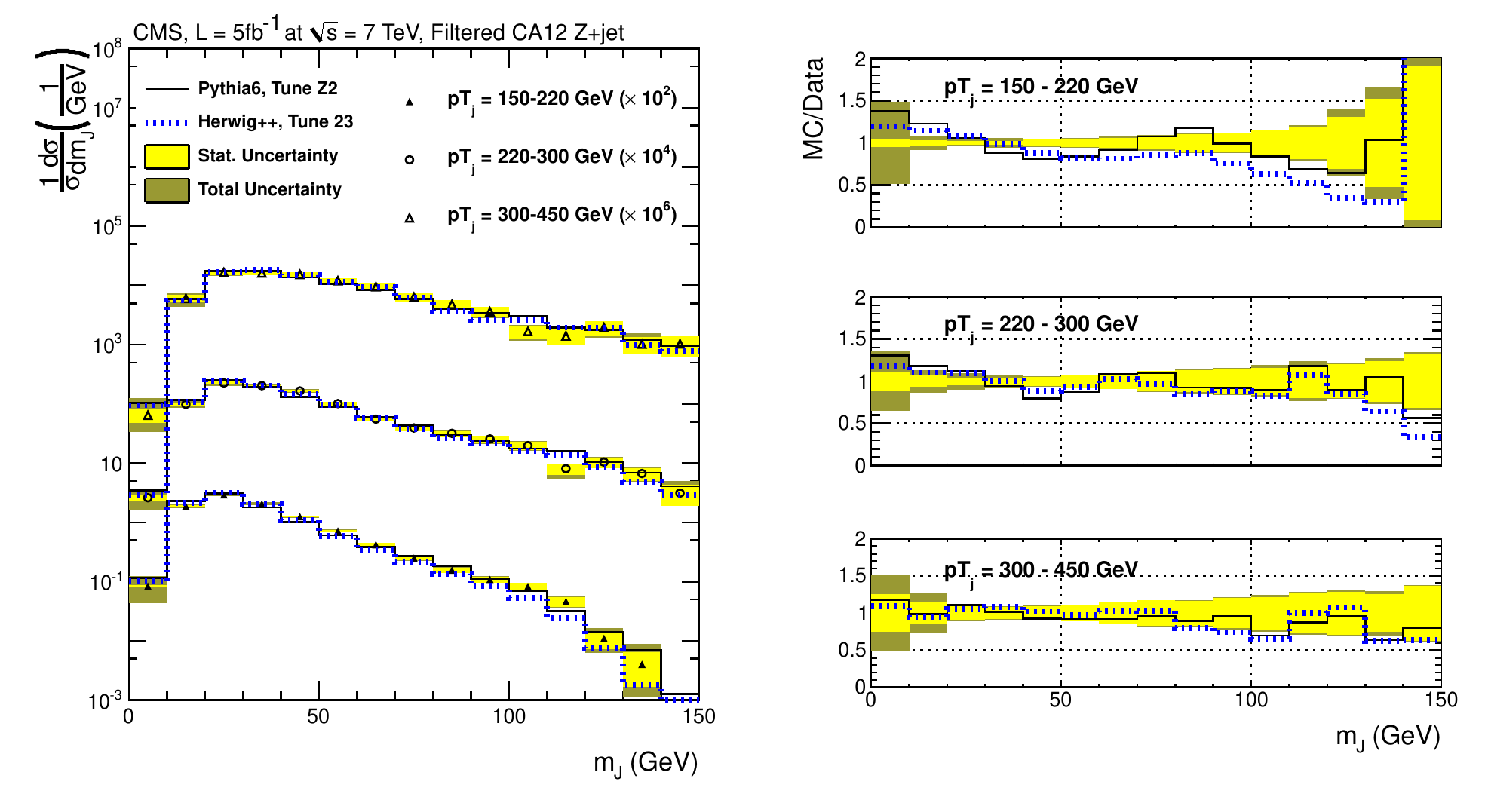}
\caption{Unfolded CA12 filtered $m_J$ distribution for $\Z(\ell\ell)$+jet events. The data (black symbols) are compared to MC expectations from {\MADGRAPH}+\PYTHIA (solid lines) and \HERWIG (dotted lines) on the left. The ratio of MC to data is given on the right.
The statistical uncertainty is shown in light shading, and the total uncertainty in dark shading.}
\label{figs:prunedZmmInt2}
\end{figure}

\begin{figure}[!htb]
\includegraphics[width=0.99\textwidth]{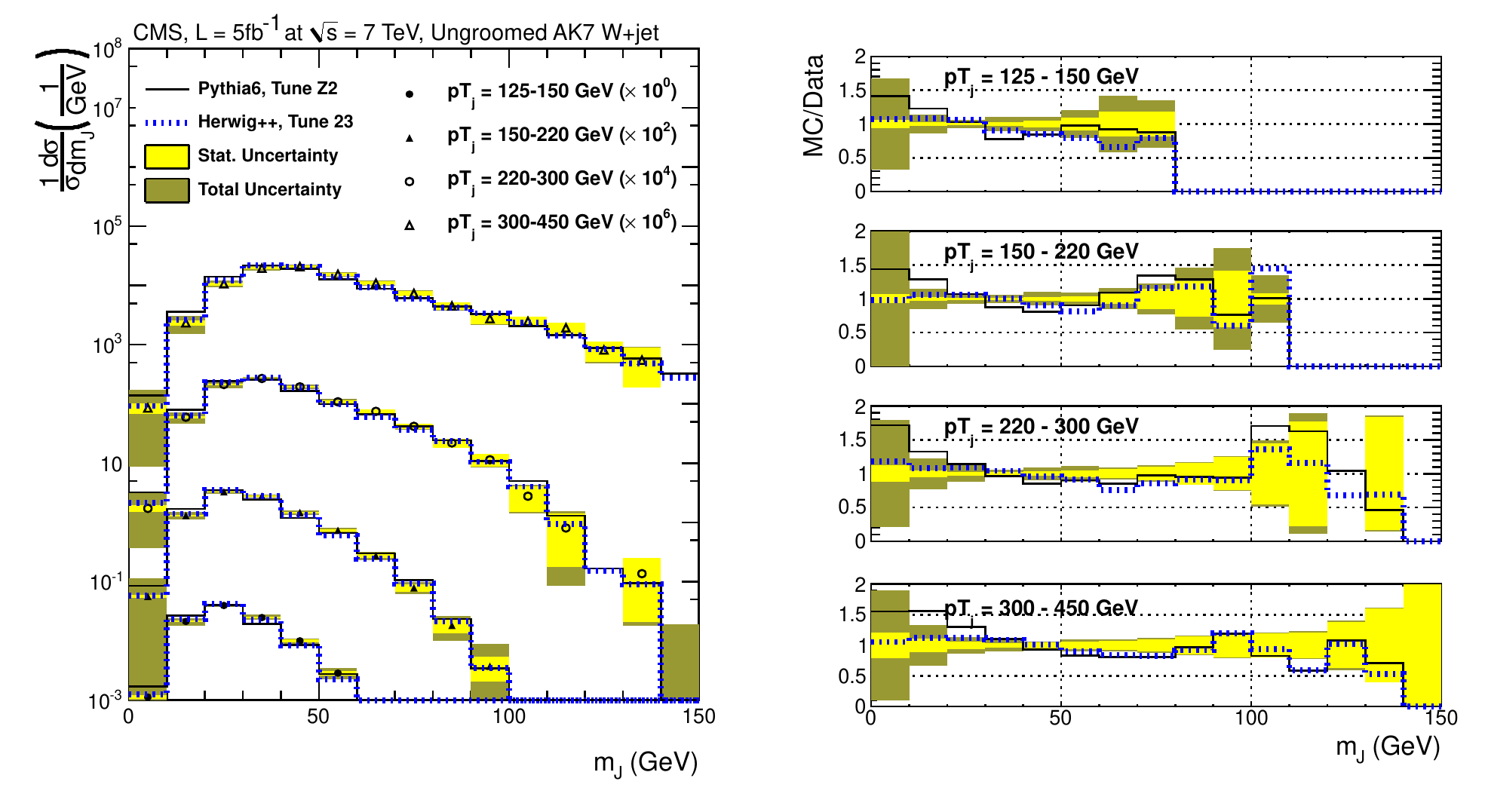}
\caption{Distributions in $m_J$ for unfolded, ungroomed AK7 jets in  \PW$(\ell\nu_\ell)$+jet events. The data (black symbols) are compared to MC expectations from {\MADGRAPH}+\PYTHIA (solid lines) and \HERWIG (dotted lines) on the left. The ratios of MC to data are given on the right.
The statistical uncertainty is shown in light shading, and the total uncertainty in dark shading.}
\label{figs:AK7WmnInt1}
\end{figure}

\begin{figure}[!htb]
\includegraphics[width=0.99\textwidth]{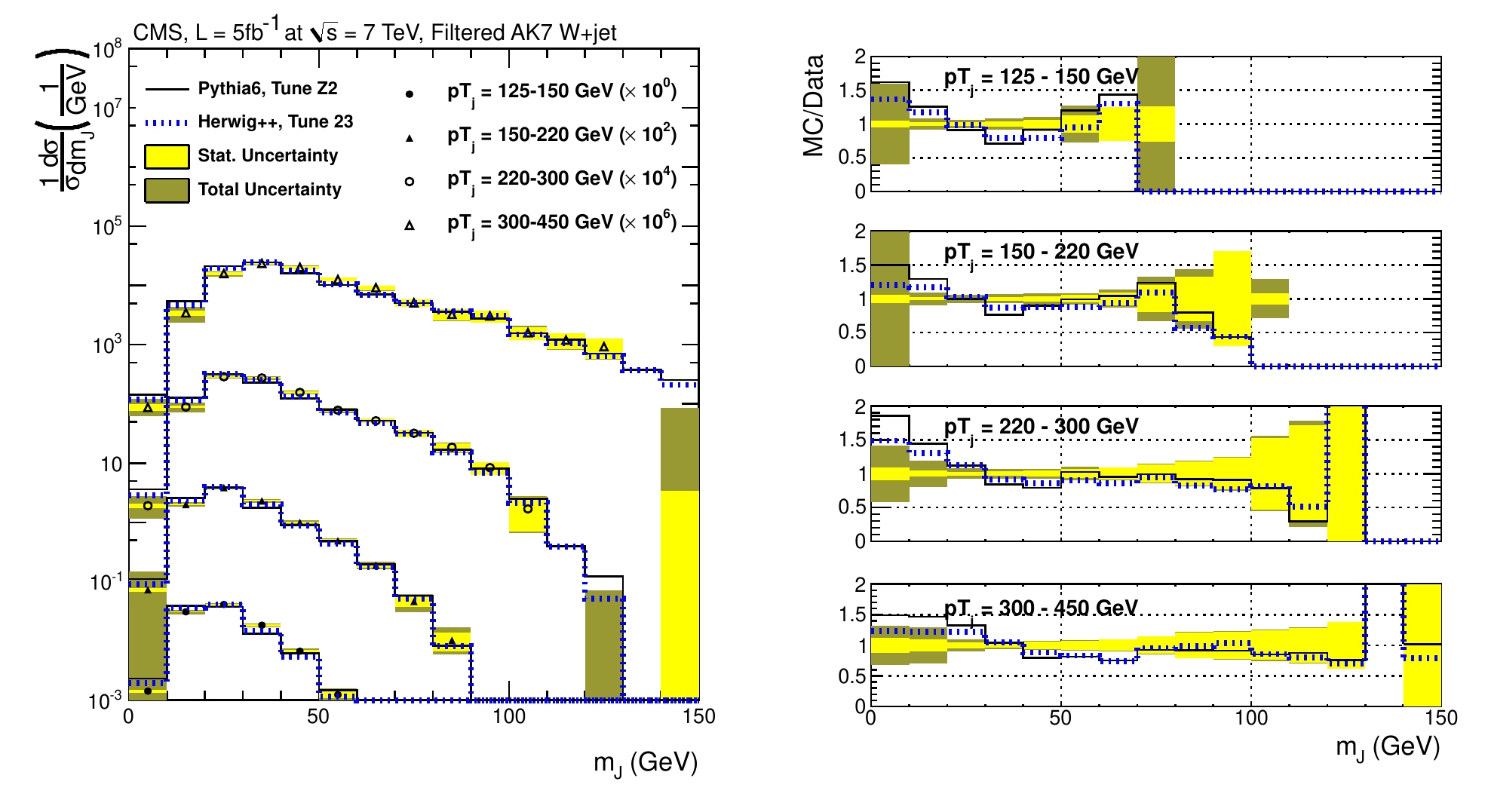}
\caption{Distributions in $m_J$ for unfolded, filtered AK7 jets in \PW$(\ell\nu_\ell)$+jet events. The data (black symbols) for different bins in $\pt$ are compared to MC expectations from {\MADGRAPH}+\PYTHIA (solid lines) and \HERWIG (dotted lines) on the left. The ratios of MC to data are given on the right.
The statistical uncertainty is shown in light shading, and the total uncertainty in dark shading.}
\label{figs:AK7WmnInt3}
\end{figure}

\begin{figure}[!htb]
\includegraphics[width=0.99\textwidth]{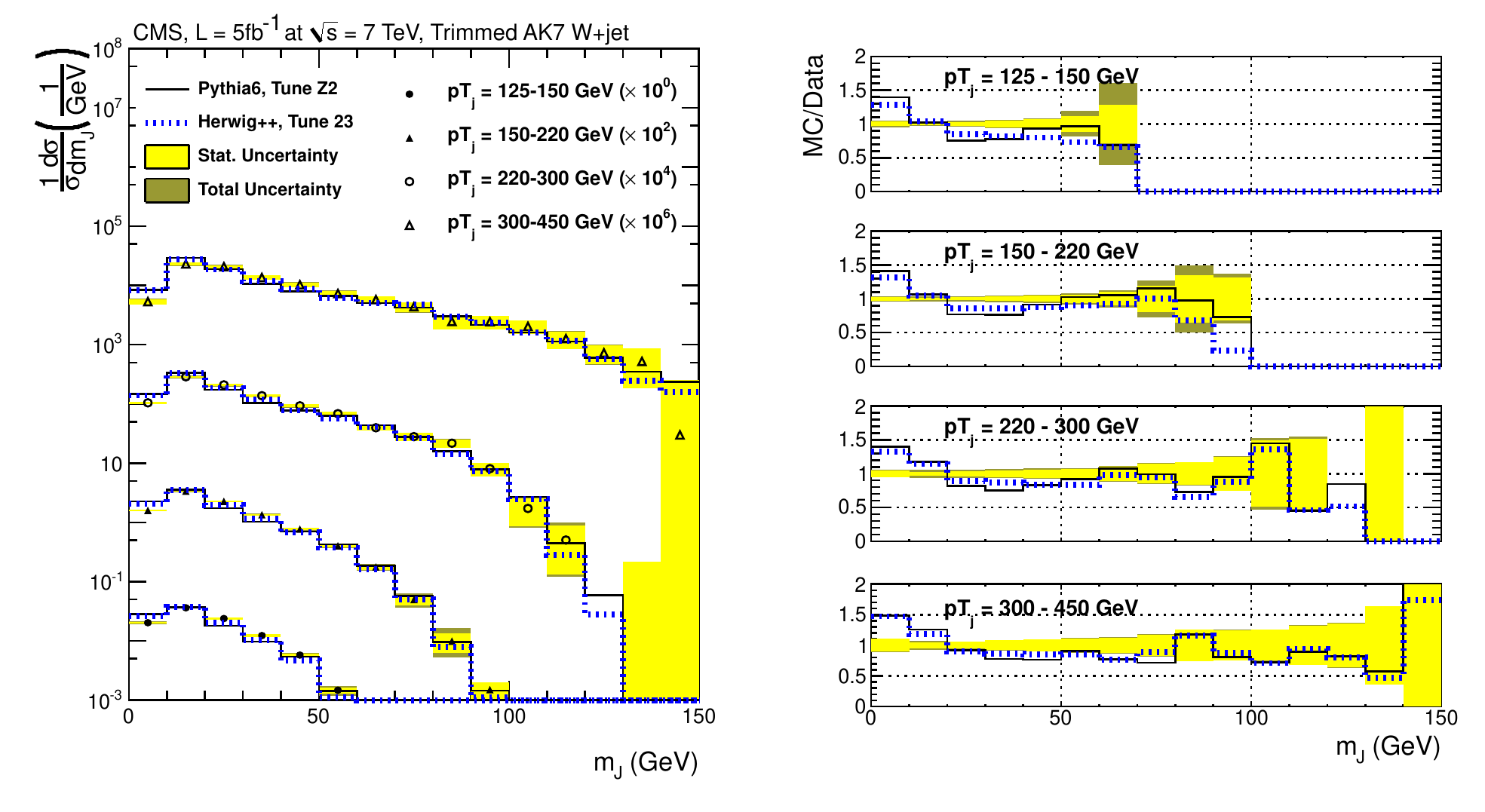}
\caption{Distributions in $m_J$ for unfolded, trimmed AK7 jets in \PW$(\ell\nu_\ell)$+jet events. The data (black symbols) for different bins in $\pt$ are compared to MC expectations from {\MADGRAPH}+\PYTHIA (solid lines) and \HERWIG (dotted lines) on the left. The ratios of MC to data are given on the right.
The statistical uncertainty is shown in light shading, and the total uncertainty in dark shading.}
\label{figs:AK7WmnInt4}
\end{figure}

\begin{figure}[!htb]
\includegraphics[width=0.99\textwidth]{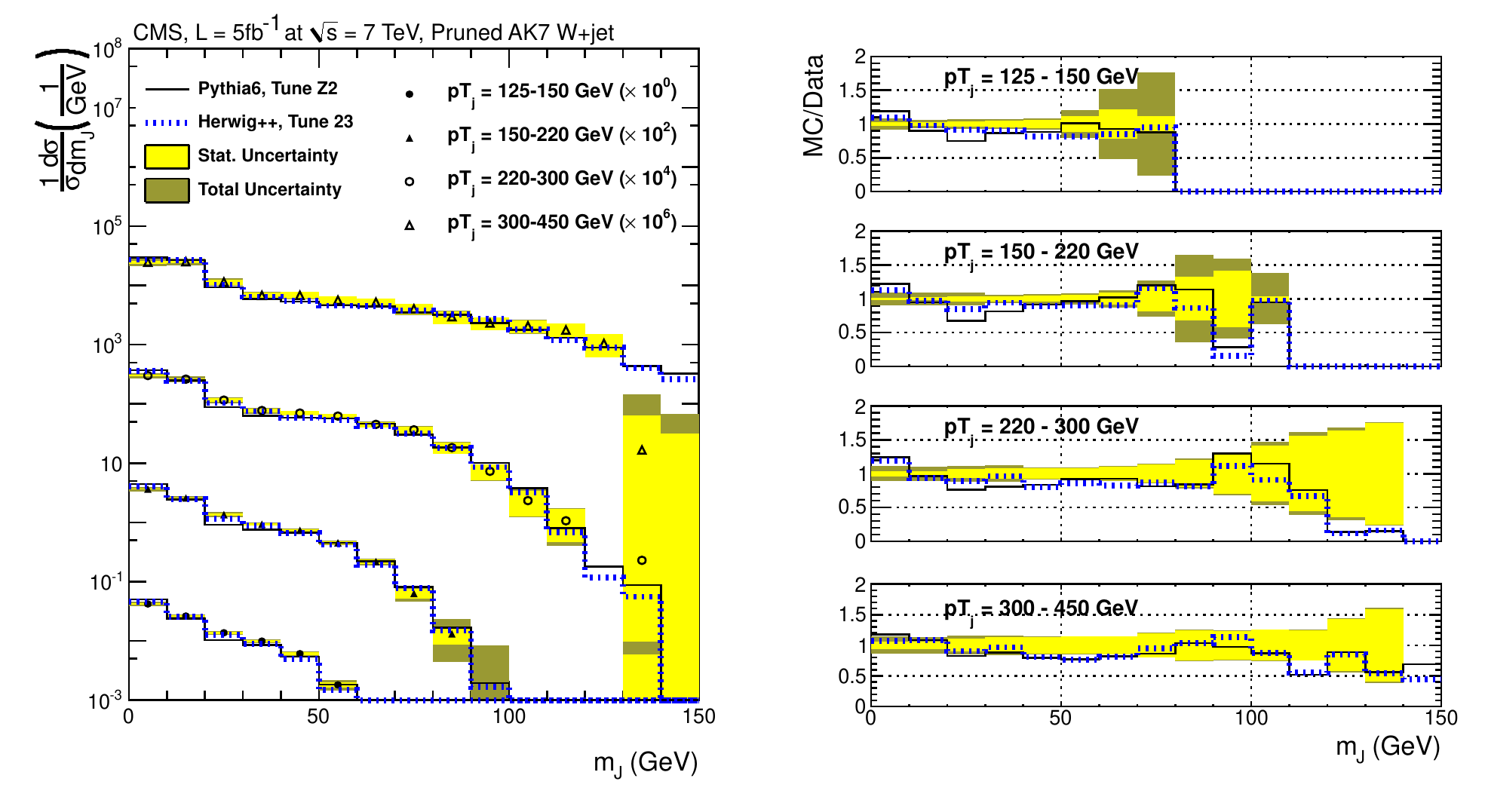}
\caption{Distributions in $m_J$ for unfolded, pruned AK7 jets in \PW$(\ell\nu_\ell)$+jet events. The data (black symbols) for different bins in $\pt$ are compared to MC expectations from {\MADGRAPH}+\PYTHIA (solid lines) and \HERWIG (dotted lines) on the left. The ratios of MC to data are given on the right.
The statistical uncertainty is shown in light shading, and the total uncertainty in dark shading.}
\label{figs:AK7WmnInt2}
\end{figure}

\begin{figure}[!htb]
\centering
\includegraphics[width=0.99\textwidth]{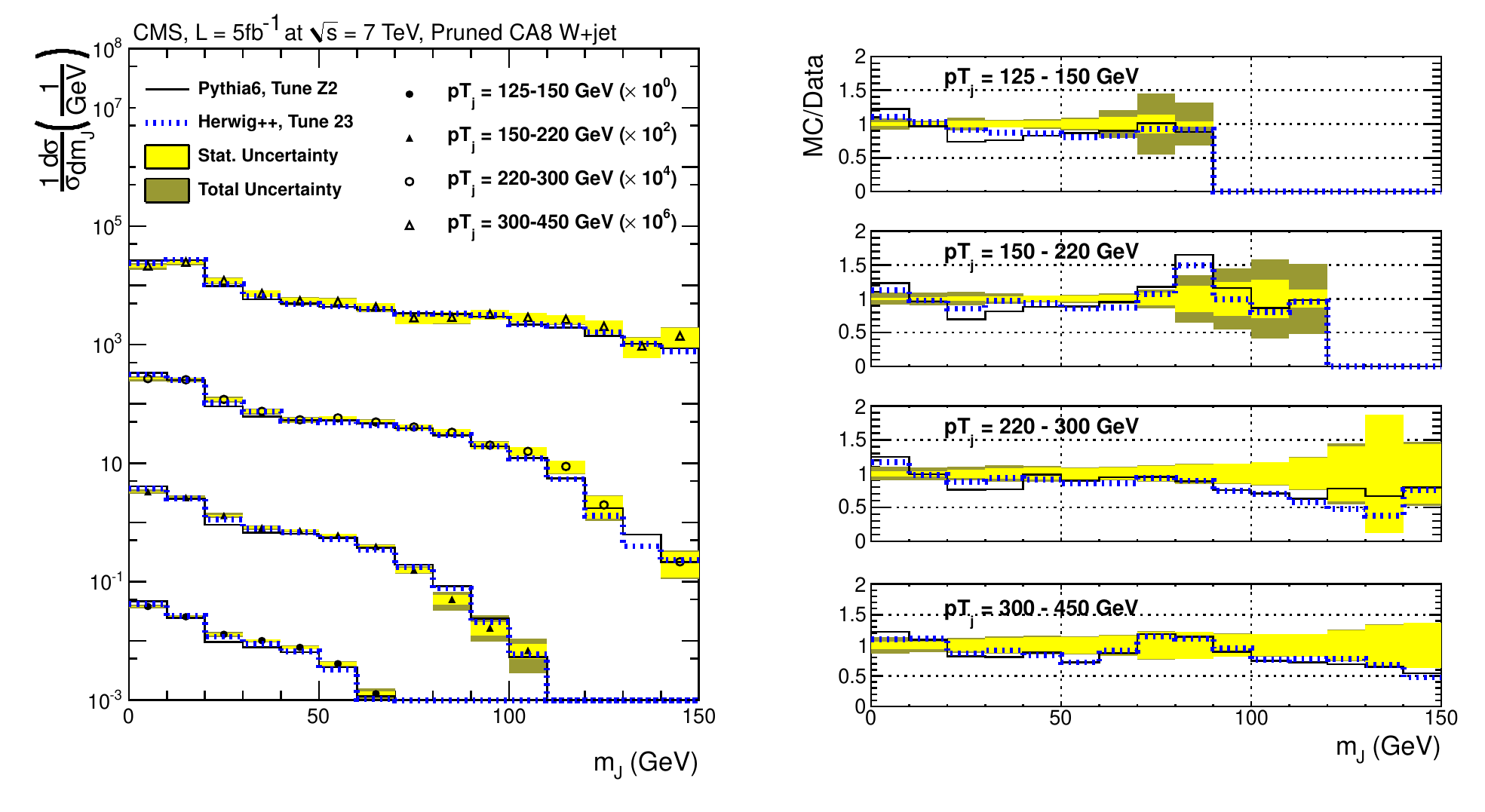}
\caption{Distributions in $m_J$ for unfolded, pruned CA8 jets in \PW$(\ell\nu_\ell)$+jet events. The data (black symbols) for different bins in $\pt$ are compared to MC expectations from {\MADGRAPH}+\PYTHIA (solid lines) and \HERWIG (dotted lines) on the left. The ratios of MC to data are given on the right.
The statistical uncertainty is shown in light shading, and the total uncertainty in dark shading.}
\label{figs:prunedWmnInt1}
\end{figure}

\begin{figure}[!htb]
\centering
\includegraphics[width=0.99\textwidth]{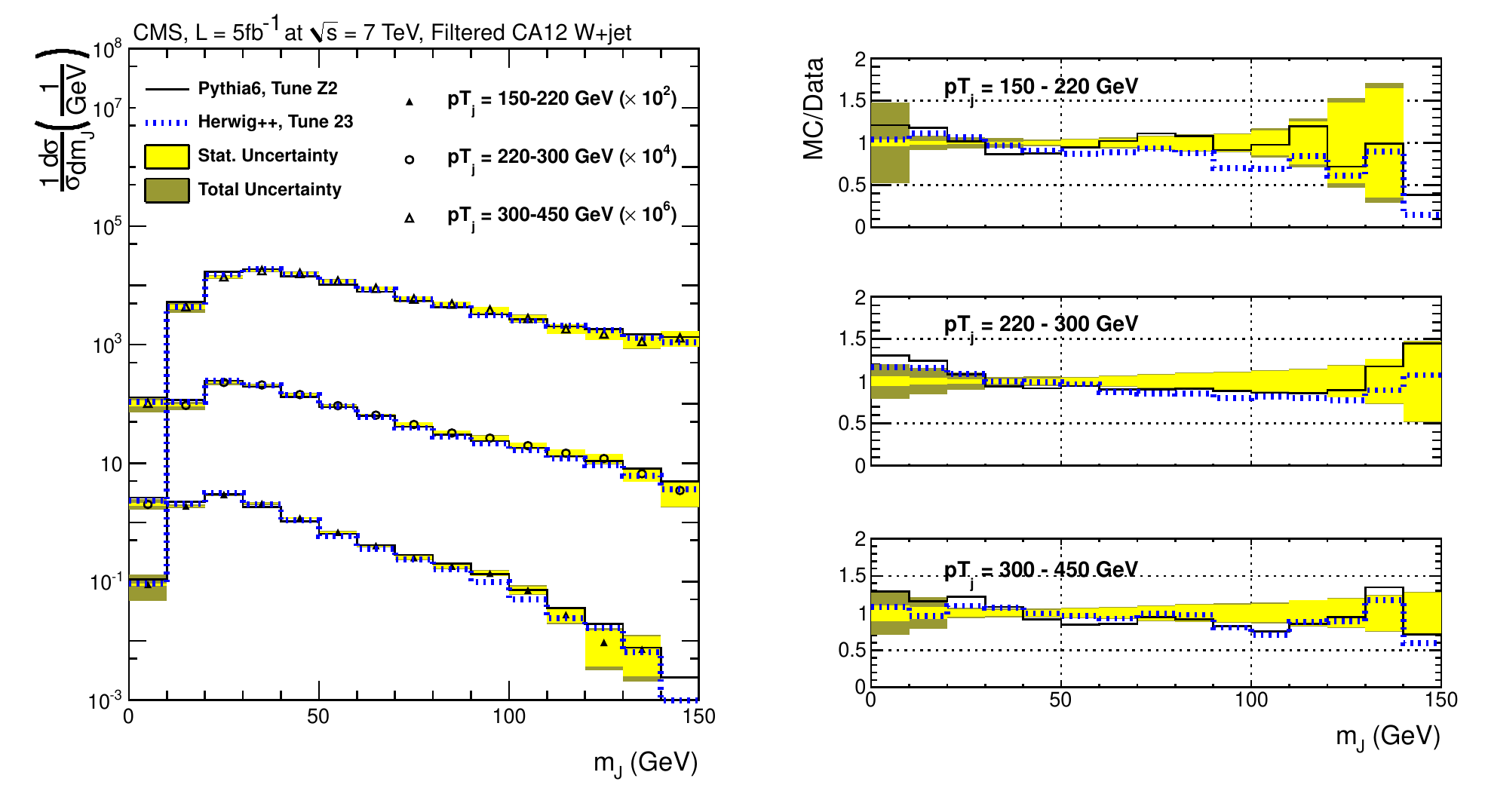}
\caption{Distributions in $m_J$ for unfolded, filtered CA12 jets in \PW$(\ell\nu_\ell)$+jet events. The data (black symbols) for different bins in $\pt$ are compared to MC expectations from {\MADGRAPH}+\PYTHIA (solid lines) and \HERWIG (dotted lines) on the left. The ratios of MC to data are given on the right.
The statistical uncertainty is shown in light shading, and the total uncertainty in dark shading.}
\label{figs:prunedWmnInt2}
\end{figure}

Figures~\ref{figs:AK7ZmmInt1}--\ref{figs:AK7ZmmInt2} show mass
distributions for the leading AK7 jet accompanying a \Z boson in
$\Z(\ell\ell)$+jet events for the ungroomed, filtered, trimmed,
and pruned clustering of jets, respectively. Both \PYTHIA and
\HERWIG show good agreement with data for all $\pt$ bins, but
especially so for $\pt > 300\GeV$. As in the case of the dijet
analysis, the data at small jet mass are not modeled satisfactorily, but
show modest improvement after applying the grooming procedures.
To investigate several popular choices of jet grooming at CMS,
Figs.~\ref{figs:prunedZmmInt1}--\ref{figs:prunedZmmInt2} show the
distributions in $m_J$ for pruned CA8 and filtered CA12 jets in {\Z}+jet
events.  For groomed CA jets, both  \PYTHIA and \HERWIG provide
good agreement with the data, with some possible inconsistency for
$m_J< 20\GeV$ and at large $m_J$ for $\pt< 300\GeV$ for the ungroomed and
filtered jets.
Figures~\ref{figs:AK7WmnInt1}--\ref{figs:AK7WmnInt2} show the
corresponding distributions for the mass of the leading jet
accompanying the \PW~boson for AK7 jets in \PW$(\ell\nu_\ell)$+jet
events for the ungroomed, filtered, trimmed, and pruned clustering
algorithms, and
Figs.~\ref{figs:prunedWmnInt1}--\ref{figs:prunedWmnInt2} show the
distributions for pruned CA8 and filtered CA12 jets.
For CA8 and CA12 jets, only particular grooming algorithms and $\pt$ bins are chosen for illustration.
The MC simulation shows good agreement with data, just as observed for
{\Z}+jet events.

\clearpage

\clearpage

\section{Summary}
\label{sec:summary}

We have presented the differential distributions in jet mass for inclusive dijet
and V+jet events,
defined through the
anti-$\kt$ algorithm for a size parameter of 0.7 for ungroomed jets, as well as
for jets groomed through filtering, trimming, and pruning. In
addition, similar distributions for V+jet events were
given for pruned Cambridge--Aachen jets with a size parameter of 0.8,
as well as for filtered Cambridge--Aachen jets with a size parameter of 1.2.
The impact of pileup on jet mass was also investigated.

Higher-order QCD matrix-element predictions for partons, coupled to
parton-shower Monte Carlo programs that generate jet mass in dijet and
V+jet events, are found to be in good agreement with data.
A comparison of data with MC simulation indicates that
both \PYTHIA and \HERWIG reproduce the data reasonably well,
and that the \HERWIG predictions for more aggressive grooming
algorithms, \ie, those that remove larger fractions of contributions
to the original ungroomed jet mass, agree somewhat better with
observations. It is also observed that the more aggressive grooming
procedures lead to somewhat better agreement between data and MC simulation.

In comparing the results from the V+jet analysis with those for the two leading jets in multijet events,
the predictions provide slightly better agreement with the
V+jet data. This observation suggests that simulation of quark jets is better than of gluon jets.
Differences between data and simulation are larger at small jet mass values, which also correspond to the region
more affected by pileup and soft QCD radiation.

These studies represent the first detailed investigations of
techniques for characterizing jet substructure based on data
collected by the CMS experiment at a center-of-mass energy of 7\TeV. For the trimming and pruning algorithms,
these studies mark the first publication on this subject from the LHC, and provide an important benchmark
for their use in searches for massive particles. Finally, the intrinsic stability of these algorithms to pileup effects
is likely to contribute to a more rapid and widespread use of these techniques in future high-luminosity
runs at the LHC.

\section*{Acknowledgements}
\hyphenation{Bundes-ministerium Forschungs-gemeinschaft Forschungs-zentren} We congratulate our colleagues in the CERN accelerator departments for the excellent performance of the LHC and thank the technical and administrative staffs at CERN and at other CMS institutes for their contributions to the success of the CMS effort. In addition, we gratefully acknowledge the computing centres and personnel of the Worldwide LHC Computing Grid for delivering so effectively the computing infrastructure essential to our analyses. Finally, we acknowledge the enduring support for the construction and operation of the LHC and the CMS detector provided by the following funding agencies: the Austrian Federal Ministry of Science and Research and the Austrian Science Fund; the Belgian Fonds de la Recherche Scientifique, and Fonds voor Wetenschappelijk Onderzoek; the Brazilian Funding Agencies (CNPq, CAPES, FAPERJ, and FAPESP); the Bulgarian Ministry of Education, Youth and Science; CERN; the Chinese Academy of Sciences, Ministry of Science and Technology, and National Natural Science Foundation of China; the Colombian Funding Agency (COLCIENCIAS); the Croatian Ministry of Science, Education and Sport; the Research Promotion Foundation, Cyprus; the Ministry of Education and Research, Recurrent financing contract SF0690030s09 and European Regional Development Fund, Estonia; the Academy of Finland, Finnish Ministry of Education and Culture, and Helsinki Institute of Physics; the Institut National de Physique Nucl\'eaire et de Physique des Particules~/~CNRS, and Commissariat \`a l'\'Energie Atomique et aux \'Energies Alternatives~/~CEA, France; the Bundesministerium f\"ur Bildung und Forschung, Deutsche Forschungsgemeinschaft, and Helmholtz-Gemeinschaft Deutscher Forschungszentren, Germany; the General Secretariat for Research and Technology, Greece; the National Scientific Research Foundation, and National Office for Research and Technology, Hungary; the Department of Atomic Energy and the Department of Science and Technology, India; the Institute for Studies in Theoretical Physics and Mathematics, Iran; the Science Foundation, Ireland; the Istituto Nazionale di Fisica Nucleare, Italy; the Korean Ministry of Education, Science and Technology and the World Class University program of NRF, Republic of Korea; the Lithuanian Academy of Sciences; the Mexican Funding Agencies (CINVESTAV, CONACYT, SEP, and UASLP-FAI); the Ministry of Science and Innovation, New Zealand; the Pakistan Atomic Energy Commission; the Ministry of Science and Higher Education and the National Science Centre, Poland; the Funda\c{c}\~ao para a Ci\^encia e a Tecnologia, Portugal; JINR (Armenia, Belarus, Georgia, Ukraine, Uzbekistan); the Ministry of Education and Science of the Russian Federation, the Federal Agency of Atomic Energy of the Russian Federation, Russian Academy of Sciences, and the Russian Foundation for Basic Research; the Ministry of Science and Technological Development of Serbia; the Secretar\'{\i}a de Estado de Investigaci\'on, Desarrollo e Innovaci\'on and Programa Consolider-Ingenio 2010, Spain; the Swiss Funding Agencies (ETH Board, ETH Zurich, PSI, SNF, UniZH, Canton Zurich, and SER); the National Science Council, Taipei; the Thailand Center of Excellence in Physics, the Institute for the Promotion of Teaching Science and Technology of Thailand and the National Science and Technology Development Agency of Thailand; the Scientific and Technical Research Council of Turkey, and Turkish Atomic Energy Authority; the Science and Technology Facilities Council, UK; the US Department of Energy, and the US National Science Foundation.

Individuals have received support from the Marie-Curie programme and the European Research Council and EPLANET (European Union); the Leventis Foundation; the A. P. Sloan Foundation; the Alexander von Humboldt Foundation; the Belgian Federal Science Policy Office; the Fonds pour la Formation \`a la Recherche dans l'Industrie et dans l'Agriculture (FRIA-Belgium); the Agentschap voor Innovatie door Wetenschap en Technologie (IWT-Belgium); the Ministry of Education, Youth and Sports (MEYS) of Czech Republic; the Council of Science and Industrial Research, India; the Compagnia di San Paolo (Torino); and the HOMING PLUS programme of Foundation for Polish Science, cofinanced from European Union, Regional Development Fund.
\bibliography{auto_generated}

\cleardoublepage \appendix\section{The CMS Collaboration \label{app:collab}}\begin{sloppypar}\hyphenpenalty=5000\widowpenalty=500\clubpenalty=5000\input{SMP-12-019-authorlist.tex}\end{sloppypar}
\end{document}

%% file: SMP-12-019-authorlist.tex
\textbf{Yerevan Physics Institute,  Yerevan,  Armenia}\\*[0pt]
S.~Chatrchyan, V.~Khachatryan, A.M.~Sirunyan, A.~Tumasyan
\vskip\cmsinstskip
\textbf{Institut f\"{u}r Hochenergiephysik der OeAW,  Wien,  Austria}\\*[0pt]
W.~Adam, E.~Aguilo, T.~Bergauer, M.~Dragicevic, J.~Er\"{o}, C.~Fabjan\cmsAuthorMark{1}, M.~Friedl, R.~Fr\"{u}hwirth\cmsAuthorMark{1}, V.M.~Ghete, N.~H\"{o}rmann, J.~Hrubec, M.~Jeitler\cmsAuthorMark{1}, W.~Kiesenhofer, V.~Kn\"{u}nz, M.~Krammer\cmsAuthorMark{1}, I.~Kr\"{a}tschmer, D.~Liko, I.~Mikulec, M.~Pernicka$^{\textrm{\dag}}$, D.~Rabady\cmsAuthorMark{2}, B.~Rahbaran, C.~Rohringer, H.~Rohringer, R.~Sch\"{o}fbeck, J.~Strauss, A.~Taurok, W.~Waltenberger, C.-E.~Wulz\cmsAuthorMark{1}
% ...
\vskip\cmsinstskip
\dag:~Deceased\\
1:~~Also at Vienna University of Technology, Vienna, Austria\\
2:~~Also at CERN, European Organization for Nuclear Research, Geneva, Switzerland\\
% ...
